\documentclass[structabstract]{aa}  
\usepackage{times}
\usepackage{natbib}
\usepackage{cite}
\usepackage{float}
\usepackage{graphicx}
\usepackage{longtable}
\usepackage{supertabular}
\usepackage{txfonts}
\begin{document}
   \title{Early-type stars in the young open cluster NGC 2244\\ and in the Mon OB2 association}

   \subtitle{I. The multiplicity of O-type stars  \thanks{Based on observations collected at Observatoire de Haute-Provence (France), San Pedro M\`artir Observatory (Mexico), La Silla Observatory (European Southern Observatory) and Asiago Observatory (Italy)}}

   \author{L. Mahy \inst{1} \and  Y. Naz\'e \inst{1}\fnmsep\thanks{Postdoctoral Researcher F.R.S.-FNRS} \and G. Rauw \inst{1}\fnmsep\thanks{Research Associate F.R.S.-FNRS} \and E. Gosset \inst{1}\fnmsep\thanks{Senior Research Associate F.R.S.-FNRS} \and M. De Becker\inst{1}\fnmsep $^{\star \star}$
           \and  H. Sana \inst{2}
          \and
          P. Eenens \inst{3}
          }

   \institute{Astrophysical Institute, Li\`ege University,
              B\^at. B5C, All\'ee du 6 Ao\^ut 17, B-4000 Li\`ege, Belgium\\
              \email{mahy@astro.ulg.ac.be}
         \and
             European Southern Observatory, Alonso de Cordova 1307, Casilla 19001, Santiago 19, Chile
	 \and
	     Departamento de Astronomia, Universidad de Guanajuato, Apartado 144, 36000 Guanajuato, GTO, Mexico
             }


 
  \abstract
  {}
   {We present the results obtained from a long-term spectroscopic campaign devoted to the multiplicity of O-type stars in the young open cluster NGC\,2244 and in the Mon OB2 association.}
   {Our spectroscopic monitoring was performed over several years, allowing us to probe different time-scales. For each star, several spectral diagnostic tools are applied, in order to search for line shifts and profile variations. We also measure the projected rotational velocity and revisit the spectral classification.}
   { In our sample, several stars were previously considered as spectroscopic binaries, though only a few scattered observations were available. Our results now reveal a more complex situation. Our study identifies two new spectroscopic binaries (HD\,46149 in NGC\,2244 and HD\,46573 in Mon\,OB2). The first object is a long-period double-lined spectroscopic binary, though the exact value of its period remains uncertain and the second object is classified as an SB1 system with a period of about 10.67 days but the time series of our observations do not enable us to derive a unique orbital solution for this system. We also classify another star as variable in radial velocity (HD\,46150) and we detect line profile variations in two rapid rotators (HD\,46056 and HD\,46485). 
   }
   { This spectroscopic investigation places a firm lower limit (17\%) on the binary fraction of O-stars in NGC\,2244 and reveals the lack of short-period O$+$OB systems in this cluster. In addition, a comparison of these new results with two other well-studied clusters (NGC\,6231 and IC\,1805) puts forward possible hints of a relation between stellar density and binarity, which could provide constraints on the theories about the formation and early evolution of hot stars.}

  \maketitle

\section{Introduction}

 O-type stars are preferentially found in young open clusters or in OB associations, suggesting these two latter environments to play a key role in massive star formation. Detailed observations could provide interesting constraints enabling to choose between the different formation scenarii proposed by theoreticians (for a summary of these, see e.g. \citealt{zy07}). To this aim, massive stars in clusters are a target of choice because they represent a homogeneous population (same distance, same age and same chemical composition).
Moreover, the study of the population of O-type stars in open clusters, and in particular of their multiplicity, is of major interest for the understanding of the physical properties of these objects. Other parameters (orbital period, eccentricity, mass ratio,...) of such multiple systems but also the binary frequency can provide a unique signature of star formation and dynamical interactions within the cluster during the very early phases of its existence. 

Investigations based upon intense spectroscopic campaigns in some clusters (e.g. NGC\,6231, \citealt{san08}; IC\,1805, \citealt{deb06}) have been performed over several years. These detailed observations lead to a serious correction of the results proposed by \citet{gar01} for O-type star rich clusters, by greatly improving our knowledge of the precise physical parameters of the binaries in these clusters. In order to extend these studies, we have undertaken a similar spectroscopic monitoring of the O-type star population in another young open cluster, NGC\,2244.

With an inferred age of about 2--3 Myr \citep{che07} and a radial velocity (RV) of 26.2$\,\pm\,$3.4 km\,s$^{-1}$ \citep{kha05}, NGC\,2244 lies inside the Rosetta nebula and forms the core of the Monoceros OB2 association. This cluster is the youngest of two or three subgroups of OB stars and stellar aggregates in this association \citep{li05}. 

Some stars in NGC\,2244 were observed by \citet{pla31} and reanalyzed by \citet{pet61}. These authors found that the majority of these objects has quite constant RVs. These results were, however, subsequently challenged by \citet{abt72} and \citet{liu89}.

From a compilation of heterogeneous data, \citet{gar01} estimated the proportion of spectroscopic binaries in NGC\,2244 to be about 50\% of the O-type stars (three out of six considered as binaries). Such an O-star binary fraction would make this cluster the third binary-rich cluster in their sample after NGC\,6231 and IC\,1805. As already mentioned above, the figures for these two clusters were significantly revised recently \citep{rau04,deb06,san08}. In addition, up to now, only two spectroscopic binaries have been detected with certainty in Mon OB2: HD\,47129 (Plaskett's star, \citealt{lin08}) and HD\,48099 (\citealt{sti96}; Mahy et al. in preparation) and both of them lie outside NGC\,2244. These conflicting results clearly call for a new, thorough analysis of the binarity in NGC\,2244.

Other studies in different wavelength domains reveal additional information on the content and overall organization of the cluster. {\it Chandra} mosaic observations of NGC\,2244 were presented by \citet{wan08}. The X-ray spectra of the OB-type stars were found to be rather soft and consistent with the standard model of small-scale shocks produced, in the inner regions of the stellar winds, by the line-driving instability \citep{fel97}. The X-ray luminosities of these stars follow the canonical $L_{\rm X}/L_{\rm bol}$ relation \citep{san06b} rather closely and reveal no evidence for extra emission due to wind-wind collisions in binary systems. \citet{wan08} found that the fainter X-ray sources are strongly concentrated around HD\,46150 (O5.5\,V) whilst no strong clustering of X-ray sources was found around the most massive component HD\,46223 (O4\,V). Following these authors, there are two possible explanations for this difference: either HD\,46223 was ejected as a result of dynamical interactions (but it does not display a strong proper motion) or it may actually be younger and would not be part of the same population as the central part of the cluster. 

The present paper, based on a long-term spectroscopic campaign, is organized as follows. Section 2 describes the observing campaign. The properties of the individual objects (spectral classification, projected rotational velocity and multiplicity) are presented in Section 3. The multiplicity and the rotational velocities of massive stars in NGC\,2244 are discussed in Section 4  while Section 5 provides a summary of our results. 


\section{Observations and data reduction}

Our team collected 207 spectra of 9 stars over 9 years using several different telescopes. This sample allows us to investigate the presence of short and long-term variations for each star.

Most of our observations were obtained with the 1.52m telescope of the Observatoire de Haute-Provence (OHP) equipped with the Aurelie spectrograph. The detector used was a thin back-illuminated CCD with 2048 $\times$ 1024 pixels of 13.5~$\mu m^{2}$. The grating \#3 (600 lines/mm) allowed us to obtain spectra with a resolving power of 10000, centered on 5700$\, \AA$ and covering wavelengths between 5500$\, \AA$ and 5920$\, \AA$. Exposure times were typically 10 to 45 minutes resulting in signal-to-noise ratios larger than 150. Such high quality observations were needed to search for weak variations or for the spectroscopic signature of faint companions. We also collected spectra with the same instrument in the blue domain (centered at 4670$\, \AA$ and covering the 4450--4900$\, \AA$ region). These data were spread over 28 days in November 2007 and over 6 days in September and October 2008. The whole dataset was reduced in a classical way (bias and flatfield corrections) using the MIDAS software developed at ESO (European Southern Observatory). The spectra were wavelength calibrated using Thorium-Argon comparison spectra obtained immediately before or after the stellar spectrum. The spectra were finally normalized to the continuum by fitting polynomials of degree 4 or 5. 

Additional spectra were obtained at Observatorio Astron\'omico Nacional of San Pedro M\`artir (SPM), in Mexico, with the 2.1m telescope equipped with Espresso. This echelle spectrograph gives 27 orders in the 3780--6950$\, \AA$ wavelength domain with a resolving power of R = 18000. The CCD detector was a SITE 3 optical chip with 1024 $\times$ 1024 pixels of 24~$\mu m^{2}$. Typical exposure times range from 5 to 15 minutes. We had to add consecutive spectra of a given night at the expense of time resolution to obtain signal-to-noise ratios close to 300. The data were reduced using the echelle package available within the MIDAS software.

In addition, we gathered spectra from both Elodie and ESO archives. Elodie was an echelle spectrograph giving 67 orders in the 3850--6850$\, \AA$ spectral range (R = 42000) mounted at the 1.93m telescope at OHP. We retrieved 13 spectra spread between November 1999 and November 2005. The mean signal-to-noise ratio for our targets was 150 for an exposure time ranging from 10 min to 1 hour. The ESO archives provided us with 11 spectra taken with FEROS (Fiber-fed Extended Range Optical Spectrograph) mounted at the ESO/MPG 2.20m telescope at La Silla (R = 48000). The data reduction was performed using a modified FEROS pipeline working under the MIDAS environment. Beyond the  modifications already described in \citet{san06a}, several new features were implemented. We used a 2D fit of the order position for improved stability. We also used the wavelength calibration frames obtained with the ``new'' ThArNe calibration lamp (see e.g. FEROS-II user manual 77.0, \S2.5). The latter however heavily saturates in the red, even on the shortest exposures planned by the calibration plan\footnote{As a reminder, the latter provides 3 series of two exposures with increasing exposure times of 3s, 15s and 30s, a scheme that is repeated at the beginning and at the end of the night.}.  Therefore, we computed a master calibration frame for each exposure time. We then proceeded with the extraction of the calibration spectra separately for each master  frame. We finally used the detected lines from the 3s/15s/30s master wavelength calibration frames to respectively calibrate the orders 30--39/10--29/1--9. The saturated lines or the lines outside the linearity range of the detector were rejected from the fit and the detection threshold was optimised for each order. Finally, a 2-iteration 3-sigma clipping method was used to discard the very few remaining discordant lines (mostly because of artifacts related to poorly corrected bad column effects). The obtained wavelength calibration residuals were all in the range 2.7-2.9~m\AA. 

At last, we retrieved two spectra from Asiago archives, with a mean signal-to-noise ratio of about 155, to increase the dataset of HD\,46149. These two spectra were taken by the 1.82m Copernicus telescope equipped with the echelle instrument (R = 28600). Unfortunately, no flatfields were available and we could only apply an overscan correction. Due to the lack of a well-suited calibration atlas, the spectra were approximatively calibrated in wavelength but this first order calibration was subsequently checked using the interstellar lines. The overview of the observations is given in Table \ref{table1}\footnote{The first and the second columns give the name of the object and the membership to the cluster or association (Ogura \& Ishida, 1981). The next columns are the epochs of the campaign, the telescope/instrument used, the observed wavelength domain, the number of spectra obtained and the time elapsed between the first and the last spectrum of the run, expressed in days.} (available electronically).

For most stars in NGC\,2244, the radial velocities (RVs) and the equivalent widths (EWs) were determined by fitting Gaussians on the line profiles, except for the two rapid rotators. Indeed, Gaussians properly fitted the observed profile of moderately rotationally broadened lines in massive stars. To measure the RVs, we fitted the bottom half of the line profiles while the entire line was used to estimate the EWs. For the two rapid rotators, we used a synthetic line profile generated with the projected rotational velocity of the star to determine the Doppler shifts and we integrated the complete line profiles to measure the EWs. We note that we used the same rest wavelengths as \citet{con77} to calculate the RVs.

Moreover, we inferred the spectral type and the luminosity class of each star by using the classification criteria of \citet{con71}, \citet{con73} and \citet{mat88,mat89} hereafter referred to as ``Conti's criteria''. To support our results, we visually compared the spectra to the atlas of \citet{wal90}. We also checked the luminosity class by comparing the computed visual absolute magnitude with the photometric calibration proposed by \citet{mar06}.  To this aim, we need a value for the distance of the cluster. \citet{mas95} derived a spectroscopic parallax to this region of 1.9$\pm$0.1 kpc, more recent photometric studies place NGC\,2244 at a distance of 1.4 to 1.7~kpc \citep{hen00} in agreement with previous estimates \citep{ogu81,per91}. In the present paper, we adopt a mean distance for NGC\,2244 of 1.55 kpc. However, we note that a change of distance by 10\% do not modify our conclusions. Finally, we compared our spectral classification with previous results from earlier studies (notably quoted in the Reed catalogue \citealt{ree05}).

\section{The O-type star population in NGC\,2244 and Mon OB2}\label{3}

Our study of the multiplicity of O-type stars is based on two wavelength domains: a blue region (4450--4900$\, \AA$) and a yellow one (5500--5920$\,\AA$). In addition to the RV and profile analysis, the former setting provides an accurate determination of the spectral classification of each star; the numerous interstellar features of the latter setting enable to check the quality of the wavelength calibration and RV measurements. We provide, in Table 2 (available electronically at CDS), the heliocentric Julian days (HJD) and the RVs of each spectrum while Table \ref{tab2} quotes, object by object, the mean RVs and the 1--$\sigma$ standard deviation of the different lines. The data are expressed in\,km\,s$^{-1}$.

In the following, an object will be considered as a true spectroscopic binary (Section \ref{3.1}) if we detect either the presence of significant, periodic RV variations or the signature of a companion whose spectral lines move in anti-phase with those of the main star.
If a star fulfills the former criterion, a full orbital solution can be computed, leading to
the precise determination of the orbital parameters. When the latter criterion is fulfilled,
it only allows us to evaluate the spectral type of the two components and their mass ratio, with only loose constraints on the orbital period. 
If none of these criteria is fulfilled, but significant RV variations are detected, then the 
star will be considered as a "RV variable", and thus a binary {\it candidate}. The RV variations are considered as significant if they are larger or equal to three times the average RV measurement error. For all stars except the two rapid rotators, this error is well represented (as was confirmed by tests on individual data) by the RV dispersion of the narrow diffuse interstellar band (DIB) at\,$\lambda$\,5780 which has similar strength and width than the stellar lines. For the rapid rotators, $\chi^2$ evaluation from shifts of the synthetic line profile showed the RV error to be about 10~km\,s$^{-1}$. Object for which significant RV variations are detected is presented in Section \ref{3.2}. The remaining objects are presented in Section \ref{3.3}.

\setcounter{table}{2}
\begin{table*}[htbp]
\caption{The mean radial velocities and the 1-$\sigma$ dispersions.}             
\label{tab2}      
\centering          
\begin{tabular}{l c c c c c c c c c}   
\hline\hline       
                      
Star & \ion{He}{i}\,$\lambda$\,4471 & \ion{He}{ii}\,$\lambda$\,4542 & \ion{He}{ii}\,$\lambda$\,4686 & \ion{O}{iii}\,$\lambda$\,5592& DIB\,$\lambda$\,5780& \ion{C}{iv}\,$\lambda$\,5801 & \ion{C}{iv}\,$\lambda$\,5812 & \ion{He}{i}\,$\lambda$\,5876& \ion{Na}{i}\,$\lambda$\,5890\\ 
\hline                    
HD\,46056 & 36.1$\,\pm\,$6.9 & 49.6$\,\pm\,$9.3 & 30.4$\,\pm\,$12.1 & \dots & 26.0$\,\pm\,$4.1& \dots & \dots & 28.7$\,\pm\,$9.7 & 23.3$\,\pm\,$2.0\\ \\
HD\,46149 & 27.3$\,\pm\,$21.8 & 27.6$\,\pm\,$18.7& 28.4$\,\pm\,$22.0& 12.6$\,\pm\,$21.4 & 23.5$\,\pm\,$4.7 & 24.1$\,\pm\,$21.3 & 20.7$\,\pm\,$20.3 & 22.3$\,\pm\,$19.4 & 21.4$\,\pm\,$3.0\\ \\
HD\,46150 & 31.2$\,\pm\,$4.0 & 45.3$\,\pm\,$8.1 & 44.8$\,\pm\,$5.0 & 26.9$\,\pm\,$7.8 &23.8$\,\pm\,$3.9& 48.8$\,\pm\,$11.7& 34.0$\,\pm\,$12.9& 32.6$\,\pm\,$5.4& 20.7$\,\pm\,$2.4\\ \\
HD\,46202 & 39.8$\,\pm\,$3.1 & 38.8$\,\pm\,$3.8& 38.0$\,\pm\,$3.2 & 31.8$\,\pm\,$1.3 & 25.4$\,\pm\,$1.5& 40.4$\,\pm\,$2.2 & 36.1$\,\pm\,$2.2 & 40.0$\,\pm\,$1.4 & 23.5$\,\pm\,$1.2\\ \\
HD\,46223 & 34.5$\,\pm\,$6.7 & 44.5$\,\pm\,$3.2 &47.0$\,\pm\,$3.3 & 31.1$\,\pm\,$7.4 & 23.8$\,\pm\,$4.6& 50.9$\,\pm\,$2.9 & 38.6$\,\pm\,$6.4 & 35.6$\,\pm\,$2.7 & 23.4$\,\pm\,$3.4\\ \\
HD\,46485 & 31.5$\,\pm\,$7.5 & 43.5$\,\pm\,$10.7 & 34.0$\,\pm\,$9.4 & \dots & 21.8$\,\pm\,$5.5 & \dots & \dots & 30.8$\,\pm\,$11.1 & 21.3$\,\pm\,$4.0\\ \\
HD\,46573 & 47.0$\,\pm\,$6.6& 48.4$\,\pm\,$10.0& 50.2$\,\pm\,$7.6& 47.9$\,\pm\,$8.5 & 27.6$\,\pm\,$3.9& 59.5$\,\pm\,$5.7 & 53.2$\,\pm\,$6.1 & 50.3$\,\pm\,$5.4 & 26.9$\,\pm\,$1.4 \\ \\
HD\,46966 & 39.2$\,\pm\,$1.4 & 44.9$\,\pm\,$2.0 & 45.3$\,\pm\,$2.0 & 38.4$\,\pm\,$2.6 &23.4$\,\pm\,$4.7& 47.7$\,\pm\,$3.0 & 43.4$\,\pm\,$3.3 & 40.8$\,\pm\,$2.4 & 20.8$\,\pm\,$2.2 \\ \\
HD\,48279 & 31.4$\,\pm\,$3.6 & 40.6$\,\pm\,$2.6 & 38.9$\,\pm\,$5.1 & 32.8$\,\pm\,$11.8& 23.1$\,\pm\,$5.0 & 43.3$\,\pm\,$6.0 & 31.9$\,\pm\,$7.4 & 39.5$\,\pm\,$2.1 & 20.8$\,\pm\,$3.3 \\ \\
\hline
\end{tabular} 
\end{table*}

\subsection{Spectroscopic Binaries}\label{3.1}

\subsubsection{HD\,46149}

The first spectroscopic study of HD\,46149 was done by \citet{pla31} who measured a mean RV of about 45.0$\,\pm\,$1.5\,km\,s$^{-1}$. \citet{pet61} confirmed, on the basis of new data, that the star might be single. They obtained a mean RV of about 38.0$\,\pm\,$2.1\,km\,s$^{-1}$. \citet{liu89} analyzed three datasets obtained over three different epochs (November 1987, February 1988 and October 1988). They found a RV dispersion of about 10.2\,km\,s$^{-1}$ and, therefore, considered the star as a potential binary. More recently, \citet{tur08} reported some velocity variabilities of HD\,46149 due to the presence of a close spectroscopic companion.

\begin{figure}[htbp]
   \centering
   \includegraphics[width=9cm, bb=36 152 551 689, clip]{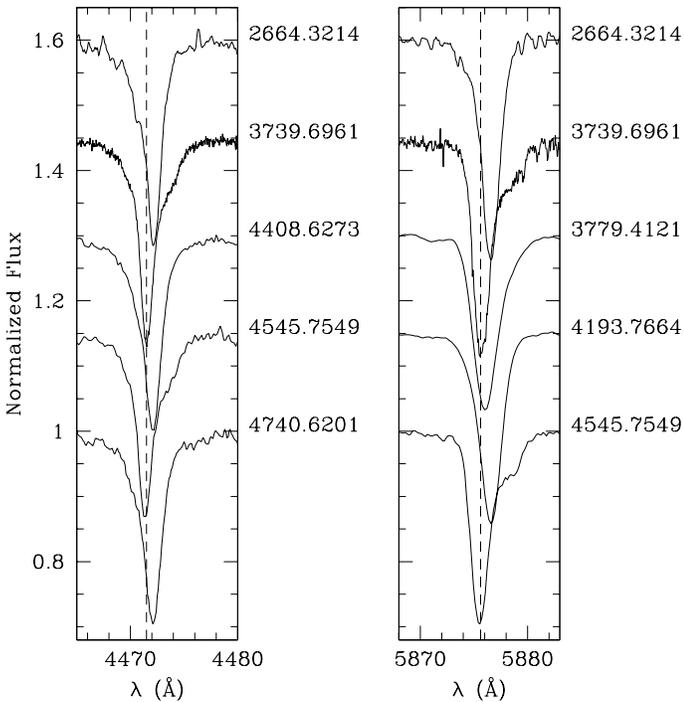}
      \caption{\ion{He}{i}\,$\lambda$\,4471 and \ion{He}{i}\,$\lambda$\,5876 line profiles of HD\,46149 observed during our spectroscopic campaigns. The time is given on the right side of the figure in the format HJD $-$ 2 450 000. All spectra are displayed in the heliocentric frame of reference. The vertical dashed line represents the rest wavelength. The signature of the secondary is clearly seen, in particular at HJD $\sim$ 2 454 545.}
         \label{Fig46149b}
   \end{figure}

In February 2006, the measurements of RVs for six spectra over six consecutive nights did not show any significant variation ($\overline{RV}\,=\,$21.6$\,\pm\,$2.1\,km\,s$^{-1}$). Moreover, no variation was detected in the line profiles. However, if we compare these RVs with the April 2007 data (HJD$\;\sim \; $2 454 199, in Fig.\,\ref{Fig46149b}), we obtain a shift close to 25\,km\,s$^{-1}$ for the \ion{He}{i}\,$\lambda$\,5876 line, and a RV shift of about 40\,km\,s$^{-1}$ for \ion{He}{i}\,$\lambda$\,4471 is measured between the data taken in January 2006 and in November 2007. Our analysis of the whole spectroscopic campaign on HD\,46149 clearly shows significant RV shifts between the observing runs. The two extreme RV values in our sampling are approximately $-5$ (Jan.\,2006, Mar.\,2008) and $+50$\,km\,s$^{-1}$ (Apr.\,2007, Nov.\,2007 and Sept.--Oct.\,2008). In addition, the FEROS spectrum and the data collected between January and April 2008 allowed us to detect the presence of a companion in the red wing of most of the \ion{He}{i} lines (visible in Fig.\,\ref{Fig46149b} at, for example, HJD$\;\sim \; $2\,453\,739 and HJD$\;\sim \; $2\,454\,545). The determination of the individual RVs enables us, in this case, to estimate the mass ratio of the two components. At the largest observed separation (at HJD$\;\sim\;$2\,454\,545), the best precision on the RVs of both stars was obtained from analyzing the \ion{He}{i}\,$\lambda$\,5876 line, which yields velocities of $-$5.8 and $+$143.3\,km\,s$^{-1}$ for the primary and the secondary, respectively. The RV of the blended spectrum, $+$52.5\,km\,s$^{-1}$ at HJD$\;\sim\;$2\,452\,664, was measured on the \ion{He}{i}\,$\lambda$\,5876 line. From these shifts, we estimate a mass ratio ($M_1/M_2$) of about 1.6. We nevertheless emphasize that the phases for which we measured the RVs might not be the extreme ones of the orbital cycle. In consequence, our mass ratio computation must be considered preliminary at best. We also estimated the flux ratio between the primary and the secondary. We measured, from the deblended \ion{He}{i}\,$\lambda$\,5876 line, an EW for the primary of 572 m$\AA\,$which we compared with the average EW quoted by Conti\,(1973) for such stars. We obtained an optical flux ratio $L_1/(L_1+L_2)$ for the primary of about 0.7. In consequence, the flux ratio $L_1/L_2$ is close to 2.3. The obtained mass and flux ratios thus suggest an early B-type for the secondary  (for comparison, from the stellar parameters listed in \citealt{mar05}, a binary system composed of an O8\,V and an O9.5\,V would have a theoretical mass ratio of $M_1/M_2\,=\,$1.3 and a flux ratio $L_1/L_2$ of about 1.9).

We then applied the Fourier analysis on the time series of primary RVs of \ion{He}{i}\,$\lambda$\,4471, \ion{He}{ii}\,$\lambda$\,4542, \ion{C}{iv}\,$\lambda$\,5812 and \ion{He}{i}\,$\lambda$\,5876 following the method described by \citet{hec85} and revised by \citet{gos01}. In all cases, we observed the highest peaks in the periodograms but its position varies according to the studied line. For example, the computed periodogram corresponding to \ion{He}{i}\,$\lambda$\,5876 yields a period of about 47 days while the one corresponding to \ion{C}{iv}\,$\lambda$\,5812 indicates a 1600 day period. We also performed a period search on the combined RV list of \ion{He}{i}\,$\lambda$\,4471 and \ion{He}{i}\,$\lambda$\,5876. Even in this case, the detected peak in the periodogram is not outstanding enough to support a particular orbital period. The data sampling is thus too poor at the present time to constrain the period of HD\,46149.

In order to separate the two components of HD\,46149, we used our disentangling program, based on the method proposed by \citet{gl06}. This consists in an iterative process which alternately uses the spectrum of one component (shifted according to its radial velocity) to remove it from the observed spectra and to calculate a mean spectrum of the other component. Furthermore, this technique allows us to compute the RVs of each star by cross-correlation even at phases for which the lines are heavily blended. We applied the disentangling algorithm in the wavelength domains 4000--4220$\, \AA$ and 4450--4730$\, \AA$. To create the RV cross-correlation mask, we used the \ion{He}{i}\,$\lambda$\,4026, \ion{Si}{iv}\,$\lambda$\,4089, \ion{He}{i}\,$\lambda$\,4120, \ion{He}{i}\,$\lambda$\,4143, \ion{He}{i}\,$\lambda$\,4471, \ion{Mg}{ii}\,$\lambda$\,4481, \ion{He}{i}\,$\lambda$\,4713 lines for the secondary and we added the \ion{Si}{iv}\,$\lambda$\,4116, \ion{He}{ii}\,$\lambda$\,4200, \ion{He}{ii}\,$\lambda$\,4542 lines for the primary. The calculated mean spectrum of the secondary star looks similar to that of a B0\,V star in the \citet{wal90} atlas. However, because of a poor sampling of the data, this result must be considered very preliminary.

HD\,46149 was classified as an O8.5\,V\,((f)) star \citep{mas95} but we revise the classification to O8 V. Note that this does not depend on the degree of blending of the spectra. With the mass ratio computed above and a theoretical mass for the primary star close to 22 $M_{\sun}$ \citep{mar05}, we estimate the secondary mass at approximately 14 $M_{\sun}$. Therefore, the system probably consists of O8 V $+$ B0--1 V stars: HD\,46149 is the first O+B binary discovered so far in NGC\,2244.

\subsubsection{HD\,46573} 

\begin{figure}[htbp]
  \centering
  \includegraphics[width=9cm, bb=27 202 590 710, clip]{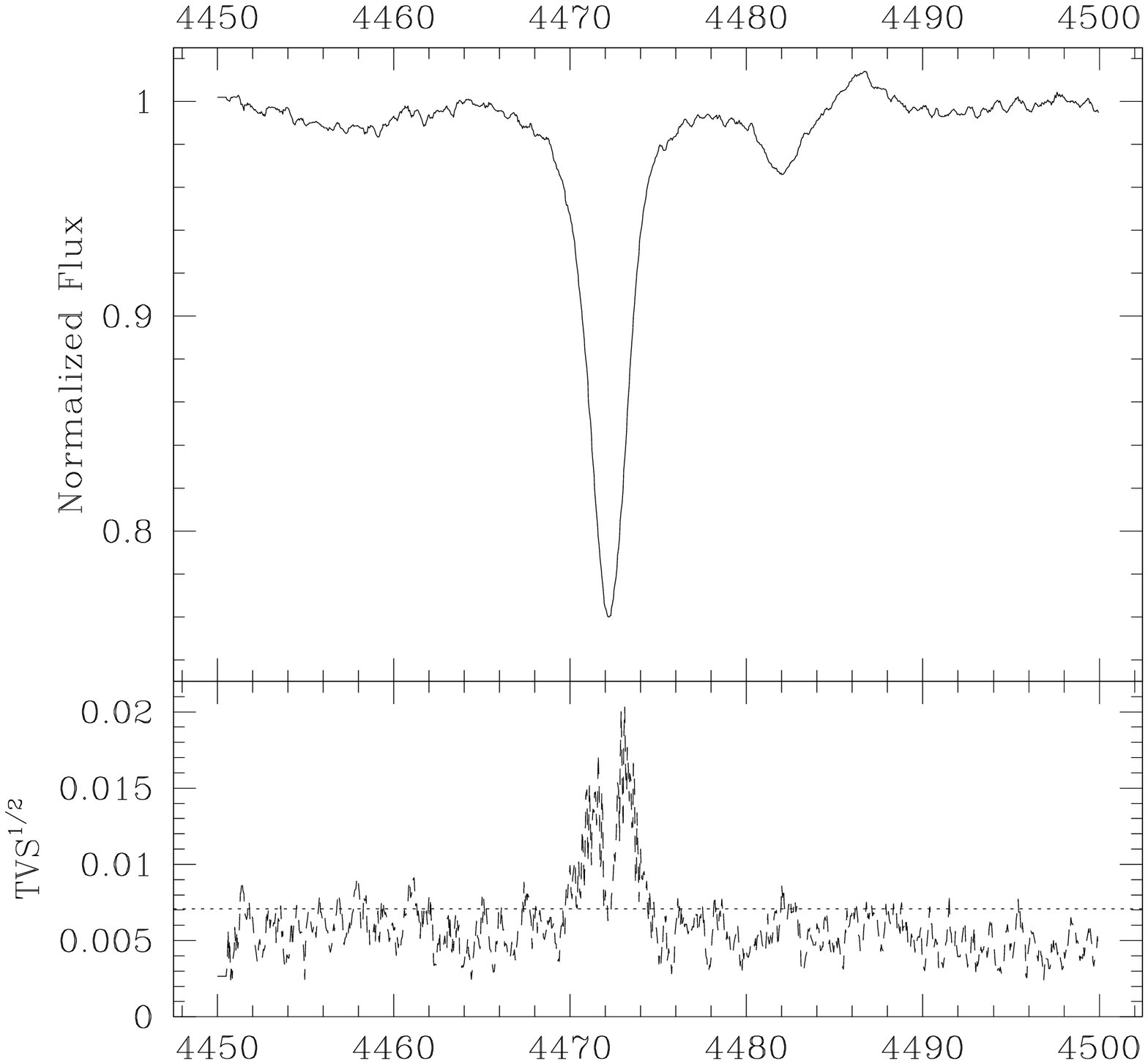}
  \includegraphics[width=9cm, bb=27 202 590 710, clip]{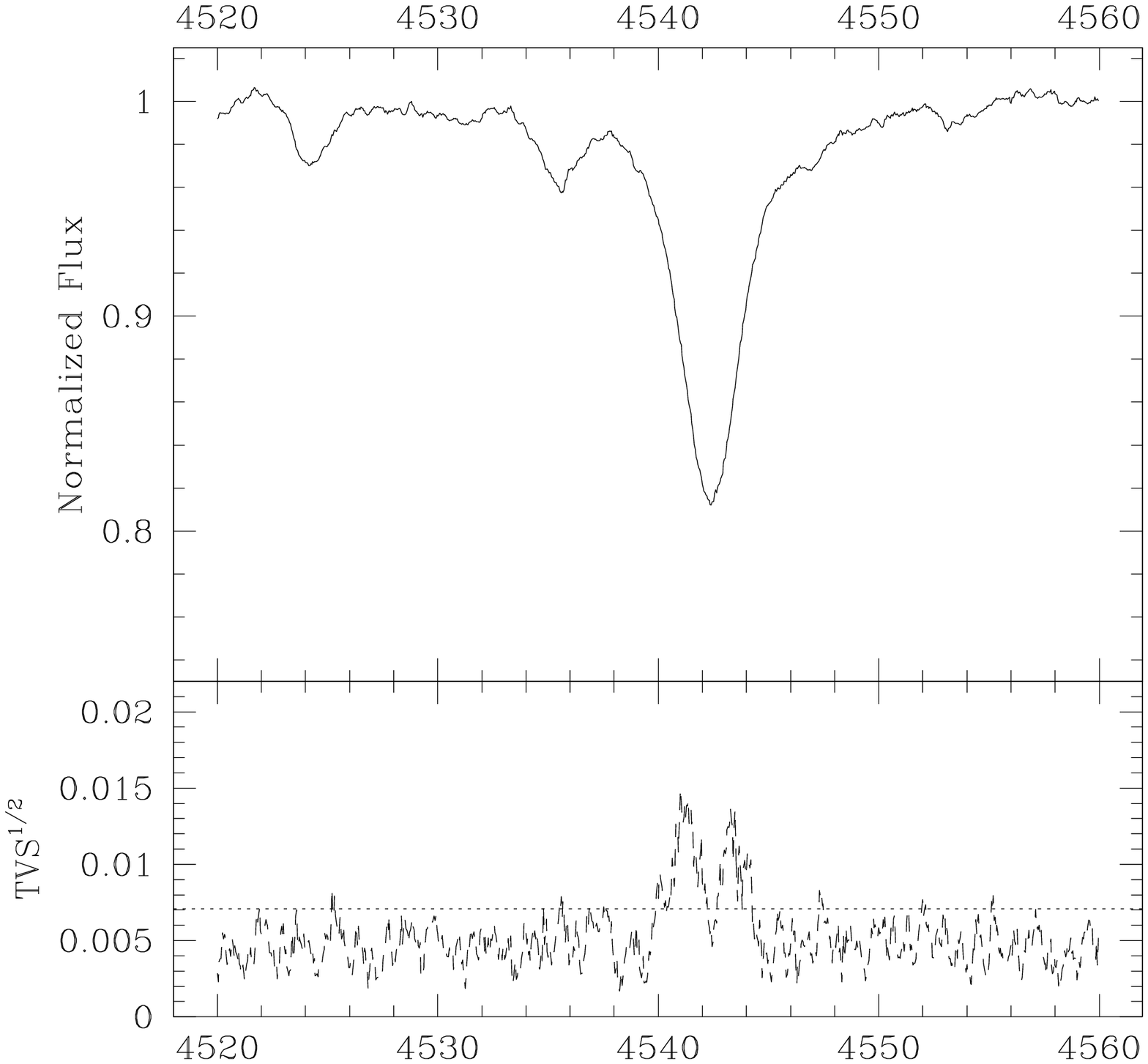}
  \caption{The mean spectrum and TVS of HD\,46573 computed from the Aurelie spectra for the \ion{He}{i}\,$\lambda$\,4471 (top) and \ion{He}{ii}\,$\lambda$\,4542 lines (bottom). In both cases, the double peaked structure is clearly visible. The dotted line illustrates the 99\% significance level for the variability evaluated following the approach of \citet{ful96}.}
  \label{Fig46573b}
\end{figure}

This star is a member of Mon OB2, but does not belong to NGC\,2244 \citep{tur76}. According to the RV variability criterion, the RV dispersions for the stellar lines of HD\,46573 are at the limit of being considered as significant. However, we computed the Temporal Variance Spectrum (TVS, \citealt{ful96}) from the Aurelie spectra for the \ion{He}{i}\,$\lambda$\,4471, \ion{He}{ii}\,$\lambda$\,4542, \ion{C}{iv}\,$\lambda \lambda$\,5801--12 and \ion{He}{i}\,$\lambda$\,5876 lines. We obtained, in each case, double peaked profiles (Fig.\ref{Fig46573b}). Together, the RV dispersion and the double peaked TVS structures lend significant support to the binary nature of the star, first envisaged by \citet{mas98}.

\begin{figure}[htbp]
  \centering
  \includegraphics[width=9cm, bb=41 168 568 708, clip]{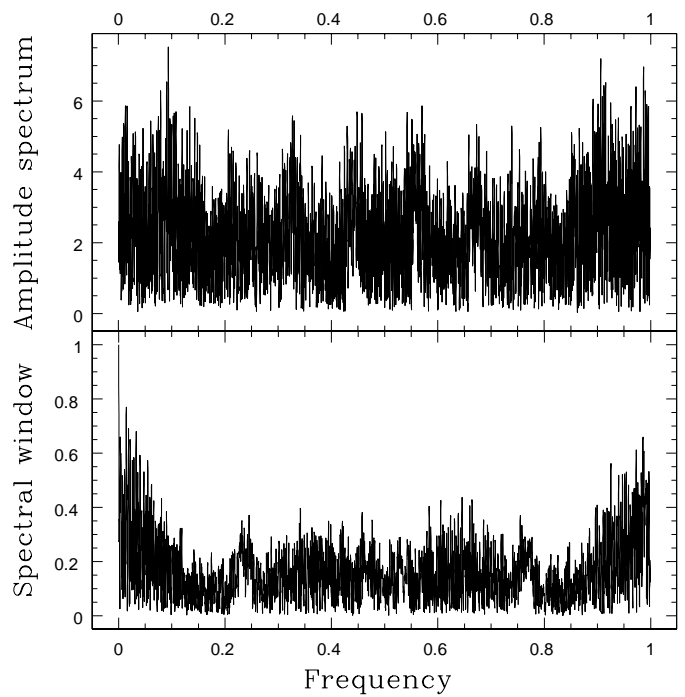}
  \caption{ The square root of the power spectrum (top) and the spectral window (bottom) of HD\,46573 computed from RVs of both \ion{He}{i}\,$\lambda$\,4471 and\,$\lambda$\,5876 lines. The y-axis corresponds to the semi-amplitude of the signal. The frequency is expressed in d$^{-1}$.}
  \label{periodogram}
\end{figure}

\begin{figure}[htbp]
  \centering
  \includegraphics[width=9cm, bb=22 156 562 692, clip]{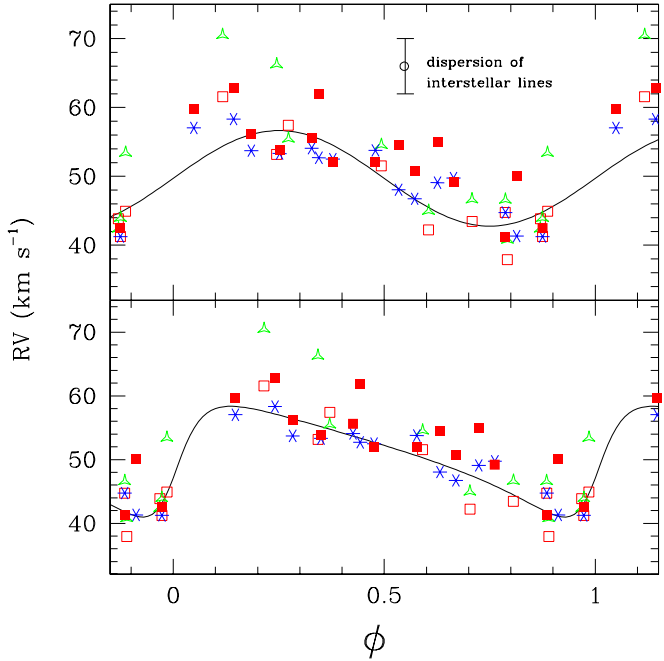}
  \caption{{\it Top: }the best-fit circular radial velocity curve for HD\,46573 computed from the RVs of the \ion{He}{i}\,$\lambda$\,4471 (open squares) and\,$\lambda$\,5876 (filled squares) lines for an orbital period of 10.67 days. The open triangles correspond to the RVs of the \ion{He}{ii}\,$\lambda$\,4542 line and the stars to the RVs of the \ion{C}{iv}\,$\lambda$\,5812 line. The error bar corresponds to $\pm$~1--$\sigma$ RV dispersion of the DIB\,$\lambda$\,5780. {\it Bottom: }the best-fit eccentric orbital solution for HD\,46573 computed for the same two lines. }
  \label{courve}
\end{figure}

We performed a Fourier analysis of the RVs. We independently applied this method to three RV time series (\ion{He}{i}\,$\lambda$\,4471, \ion{He}{ii}\,$\lambda$\,4542 and \ion{He}{i}\,$\lambda$\,5876) but the semi-amplitude spectra did not allow us to identify any dominant peak likely to be associated with a well-determined period. We then combined the RVs of the \ion{He}{i}\,$\lambda$\,4471 and \ion{He}{i}\,$\lambda$\,5876 lines. The highest peak in the resulting semi-amplitude spectrum occurs at a frequency of 0.0937\,d$^{-1}$\,(Fig.\,\ref{periodogram}) which corresponds to a period of about 10.67\,days but this value is still very preliminary since random fluctuations can not be totally excluded. The SB1 orbital solution of HD\,46573 is calculated with the program LOSP (Li\`ege Orbital Solution Package, Sana\,\&\,Gosset, A\&A submitted) based on the method of \citet{wol67}. We assigned the same weight, i.e. 1.0, to all spectra and fixed the period value to 10.67\,days. Table \ref{tabcourve} gives the main orbital elements computed for HD\,46573. $T_0$ (expressed in HJD$-$2\,450\,000) refers to the time of periastron passage for $e$\,$\neq$\,0 and to the conjunction with the primary star in front for $e$\,=\,0. $\gamma$, $K$ and $a\:\sin i$ are respectively the
systemic velocity, the semi-amplitude of the radial velocity curve, and the projected
separation between the centre of the star and the centre of mass of
the binary system. The orbital elements are given respectively for a circular and for an eccentric orbit from the RVs of both \ion{He}{i}\,$\lambda$\,4471 and\,$\lambda$\,5876. Folded with this period, our data favored a non-zero eccentricity (Fig.\,\ref{courve}, bottom panel). While not in sufficient number to calculate a full orbital solution, the RVs from \ion{He}{ii} and \ion{C}{iv} lines support the results from the \ion{He}{i} lines (Fig.\,\ref{courve}). However, there are still some phase intervals without observations and we therefore need more data to secure the period and the orbital parameters.

We then computed, with the Fourier transform method \citep{sim07}, the projected rotational velocity of HD\,46573. Previously estimated at about 95\,km\,s$^{-1}$ \citep{coe77}, we measured, for four lines (\ion{He}{ii}\,$\lambda$\,4686, \ion{O}{iii}\,$\lambda$\,5592, \ion{C}{iv}\,$\lambda$\,5801 and \ion{He}{i}\,$\lambda$\,5876), a mean value of about $v\,\sin{i}\,=\,$110$\,\pm\,$18\,km\,s$^{-1}$. 

\setcounter{table}{3}
\begin{table}[htbp]
\caption{Possible orbital parameters of the SB1 solution of HD\,46573.}    
\label{tabcourve}      
\centering          
\begin{tabular}{l c c}   
\hline\hline       
   Parameter & Circular  & Eccentric\\
\hline
P(days) & 10.67 (fixed) & 10.67 (fixed)\\
e & \dots & 0.47$\pm$0.13\\
$\omega$ (degrees)& \dots & 254.63$\pm$14.2\\
$T_0$  & 3732.24$\pm$0.26 & 3731.68$\pm$0.48\\
$\gamma$ (km s$^{-1}$)& 49.7$\pm$ 0.7 & 50.9$\pm$0.8\\
$K$ (km s$^{-1}$)&  6.9$\pm$0.9  &  8.5$\pm$1.1\\
$a\:\sin i$ ($R_\odot$) &    1.46$\pm$0.19 & 1.58$\pm$ 0.24\\
$f(m)$ ($M_\odot$) & 0.0003$\pm$ 0.0001 & 0.0005$\pm$0.0002\\
r.m.s. (km s$^{-1}$) & 3.5 & 2.6 \\
Prob ($\chi^2\,>$)&78\%&99\%\\
\hline                  
\end{tabular} 
\end{table}

The spectral type was previously quoted as O7.5\,V\,((f)) \citep{con74}, O7\,V \citep{bis82} or O7\,III\,((f)) \citep{wal71}. In our data, the \ion{He}{i}\,$\lambda$\,4471 to \ion{He}{ii}\,$\lambda$\,4542 ratio implies an O7.5 type while the EW ratio between \ion{Si}{iv}\,$\lambda$\,4088 and \ion{He}{i}\,$\lambda$\,4143 favors a Main-Sequence (MS) luminosity class. The visual absolute magnitude is more uncertain since the distance to the star is not exactly known. Assuming that Mon OB2 lies at the same distance as NGC\,2244,  we estimated $M_V$ for the two recent extreme values of the distance of NGC\,2244 (1.4 kpc and 1.7 kpc). We obtain $M_V\,=\,-4.67$ and $M_V\,=\,-5.07$, using $V\,=\,7.933$ and $E(B-V)\,=\,0.61$ \citep{mai04}. The reported results are closer to a MS star than to a giant \citep{mar06}. The spectrum also shows weak \ion{N}{iii}\,$\lambda \lambda$\,4634$-$41 emission lines and a strong \ion{He}{ii}\,$\lambda$\,4686 absorption. In H$\alpha$, we detect a weak emission in the absorption line probably due to the surrounding nebula. Thus, we favor an O7.5\,V\,((f)) spectral type, for HD\,46573. 

In summary, it exists two possible explanations for which the secondary component is not detected: a low-mass companion or a small inclination of the orbital plan. However, with such a large $v\,\sin{i}$ for the primary, it seems to be unlikely that the system is seen under a low inclination angle. In consequence, we assume that HD\,46573 is an O7.5 V\,((f)) star which possesses a lower mass companion.

\subsection{Radial velocity variables}\label{3.2}

\subsubsection{HD\,46150}\label{section33}

Reported as the second hottest star in NGC\,2244, HD\,46150 is also one of the most studied objects in this cluster. This star was first observed by \citet{pla31} who reported constant RVs. Since the 1970's, many other investigations were devoted to HD\,46150 \citep{abt72,coe77,gar80} but the interpretations of the multiplicity of the star greatly varied from one author to another. \citet{gar80} considered the changes in the RVs to be a consequence of motions in the stellar atmosphere. \citet{liu89} and \citet{und90} suggested that the star could be a spectroscopic binary even though the signature of the secondary component had not been detected yet. \citet{ful90} observed asymmetric line profiles, typical of an SB2.

\begin{figure}[htbp]
   \centering
   \includegraphics[width=9cm, bb=24 152 580 700, clip]{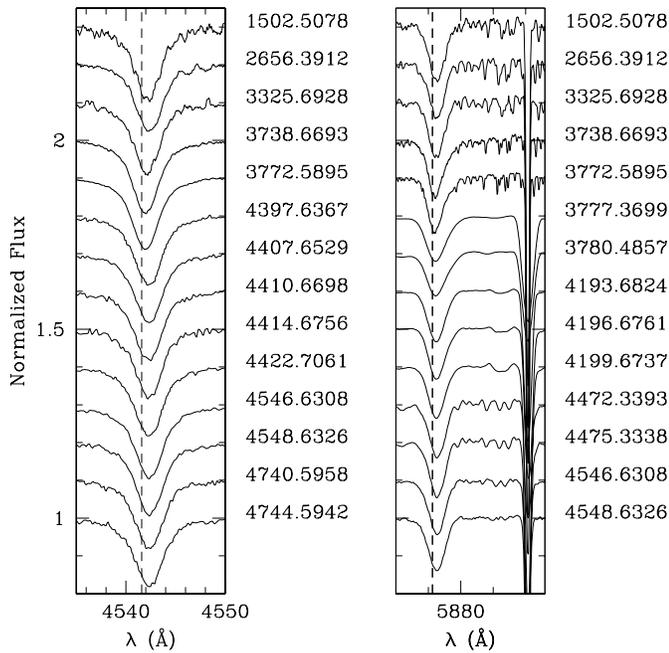}
      \caption{\ion{He}{ii}\,$\lambda$\,4542 and \ion{He}{i}\,$\lambda$\,5876 line profiles of HD\,46150 observed during our spectroscopic campaigns. }
         \label{Fig46150b}
   \end{figure}

Between November 1999 and October 2008, we have collected 34 spectra spread over different time-scales. We observed a slight RV variability in the absorptions (Fig.\,\ref{Fig46150b} notably compares the \ion{He}{ii}\,$\lambda$\,4542 line at HJD\,$\sim$\,2\,453\,772 and HJD\,$\sim$\,2\,454\,397) as well as in the emission lines which is supported by the larger dispersion for stellar lines compared to the interstellar ones. For instance, we noted a RV shift of \ion{He}{ii}\,$\lambda$\,4542 in the blue data by 28\,km s$^{-1}$ and of \ion{He}{i}\,$\lambda$\,5876 in the yellow spectra by 21\,km s$^{-1}$. Moreover, the RV dispersion is smaller for the \ion{He}{i} lines than for the \ion{He}{ii} and metallic lines which are clearly RV variable. The presence of an unresolved companion could explain this effect. Indeed, if the secondary is cooler, the \ion{He}{i} lines of the primary will be more affected by blends with the lines of the secondary component. The RV variations of lines associated with a stronger ionization (\ion{He}{ii}, \ion{C}{iv},...) will thus reflect better the true motion of the primary which leads to a larger RV dispersion. 

\begin{figure}[htbp]
  \centering
  \includegraphics[width=9cm, bb=27 200 580 720, clip]{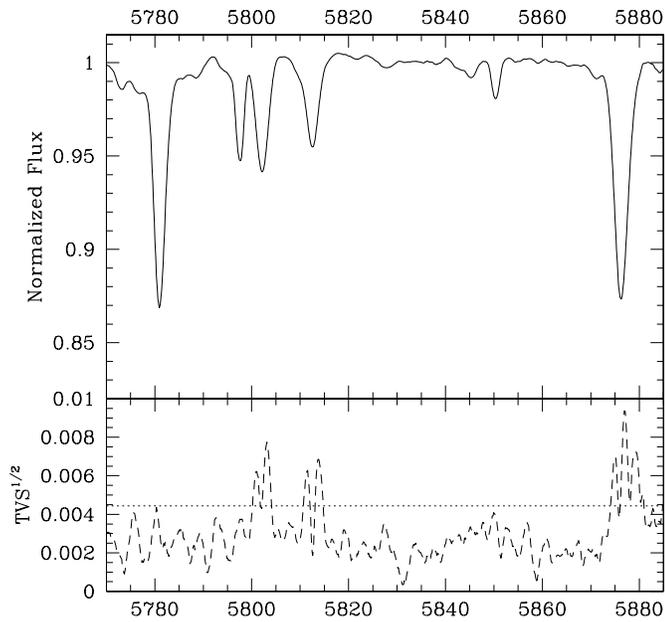}
  \caption{Mean spectrum and TVS of HD46150 for the \ion{C}{iv}\,$\lambda \lambda$\,5801$-$12 and \ion{He}{i}\,$\lambda$\,5876 lines. The dotted line illustrates the 99\% significance level for the variability evaluated following the approach of \citet{ful96}. We clearly see the double peaked structure in the stellar lines.}
  \label{Fig46150c}
\end{figure}

In order to compute the TVS from all spectra of HD\,46150, we had to degrade the spectral resolution of the echelle spectra down to that of the Aurelie data. We proceeded by convolving the echelle spectra with Gaussians to obtain a homogeneous spectral resolution for all spectra. Variations in double peaks (in the \ion{He}{i}\,$\lambda$\,4471, \ion{He}{ii}\,$\lambda$\,4542, \ion{He}{ii}\,$\lambda$\,4686 and \ion{C}{iv}\,$\lambda \lambda$\,5801--12 lines) and even variations in triple peaks, shown in Fig.\,\ref{Fig46150c}, for the \ion{He}{i}\,$\lambda$\,5876 line, are observed. Such features are reminiscent of what is usually found in spectroscopic binaries. However, no obvious signature of a companion could be found. Moreover, we note that these variations in double peaks are not visible on short time-scales ($\sim$\,7\,days), suggesting a rather long-period binary.

To check for the presence of a putative period, we applied the Fourier method to the measured RVs. We independently used this technique for the \ion{He}{i}\,$\lambda$\,4471, \ion{He}{ii}\,$\lambda$\,4542, \ion{C}{iv}\,$\lambda$\,5812 and \ion{He}{i}\,$\lambda$\,5876 lines. For each line, several peaks are present in the periodogram but their semi-amplitude nearly corresponds to the standard deviation for the RVs of the interstellar lines. We also computed the power spectrum by combining the RVs of the \ion{He}{i}\,$\lambda$\,4471 and the \ion{He}{i}\,$\lambda$\,5876 lines but, once again, the peak is not significant enough to ascertain the period.

Previously estimated to 118$\,\pm\,$25\,km\,s$^{-1}$ \citep{coe77} and 86\,km\,s$^{-1}$ \citep{pen96}, the projected rotational velocity of HD\,46150 has been determined using the Fourier method on four different lines (\ion{He}{ii}\,$\lambda$\,4686, \ion{O}{iii}\,$\lambda$\,5592, \ion{C}{iv}\,$\lambda$\,5801 and \ion{He}{i}\,$\lambda$\,5876). We find a mean value of $v\,\sin{i}\,=\,$97$\,\pm\,$9\,km\,s$^{-1}$.

In the literature, the spectral type is quoted between O5\,V\,((f)) \citep{und90,mai04} and O5.5\,V\,((f)) \citep{coe77,gar80}. From our measurements of the EWs, we confirm the latter spectral type. We observe weak \ion{N}{iii}\,$\lambda \lambda$\,4634$-$41 emission lines and a strong \ion{He}{ii}\,$\lambda$\,4686 absorption line which confirms the addition of the ((f)) suffix. We compute $M_V\,=\,-5.45\,\pm$ 0.22, from $V\,=\,6.74$ and $E(B-V)\,=\,0.40$, which corresponds to a MS star although we note that $M_V$ is somewhat brighter than the theoretical value ($M_V$\hspace{-0.15cm}\textsuperscript{theo}$\,=\,-5.12$) of \citet{mar06} for typical O5.5\,V stars. The difference could result from the approximative distance taken at 1.55 kpc or could also reflect the presence of a companion. We conclude that the spectral type of HD\,46150 is O5.5\,V\,((f)). Despite the lack of a clear periodicity or of a clear detection of the companion, several hints (double peaked structures in TVS, large $\Delta RV$,...) lead us to consider HD\,46150 as a potential binary.

\subsection{Presumably single stars}\label{3.3}

\subsubsection{HD\,46056}

\begin{figure}[htbp]
  \centering
  \includegraphics[width=7cm , bb=110 156 430 696, clip]{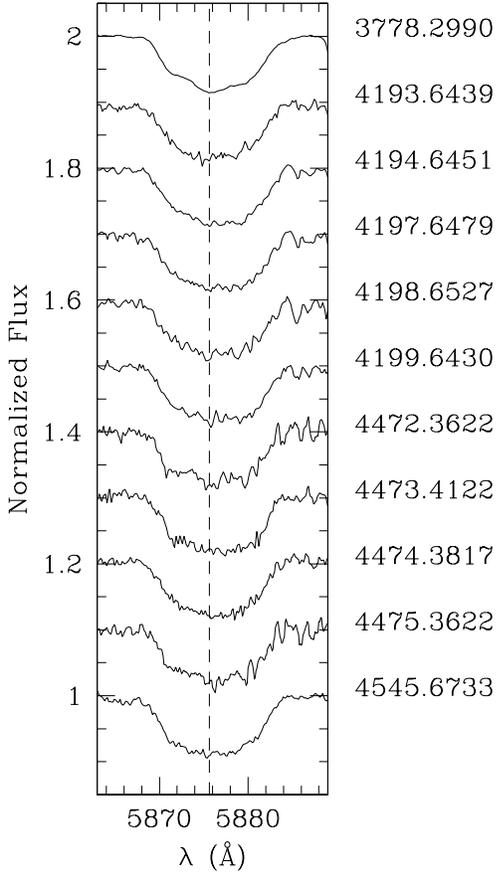}
  \caption{\ion{He}{i}\,$\lambda$\,5876 line profiles of HD\,46056 observed from February 2006 to March 2008.}
  \label{Fig46056}
\end{figure}

In the past, HD\,46056 was generally considered as a spectroscopic binary because of a variable RV \citep{wal73,liu89,und90}.
The line profiles always appear broad and shallow, suggesting the star to be a single rapid rotator rather than a spectroscopic binary. Indeed, the Helium lines display very broad profiles extending over more than 15~\AA~with a clearly non-Gaussian shape (see Fig.\,\ref{Fig46056}). Furthermore, some metallic lines, such as \ion{Mg}{ii}\,$\lambda$\,4481 or \ion{O}{iii}\,$\lambda$\,5592, are too shallow to allow a measurement of their RVs by Gaussian fits or are too heavily blended to be distinguished as it is the case for the \ion{C}{iv}\,$\lambda \lambda$\,5801--12 lines. As a consequence, we adopted a different technique to measure the RVs. We first compute the Fourier transform of different Helium line profiles using our best spectrum of the star which combines a high signal-to-noise ratio and a high dispersion. We then found the mean projected rotational velocity ($v\,\sin{i}$) to be 355$\,\pm\,$21\,km\,s$^{-1}$.

Synthetic rotation profiles with $v\,\sin{i}$\,=\,346, 350, 358 and 365\,km\,s$^{-1}$ were subsequently used to fit the \ion{He}{i}\,$\lambda$\,4471, \ion{He}{ii}\,$\lambda$\,4542, \ion{He}{ii}\,$\lambda$\,4686 and \ion{He}{i}\,$\lambda$\,5876 lines, respectively. For all these lines, the mean of the differences between the two extreme RV values were $28.1$\,km\,s$^{-1}$ and the mean 1$-\sigma$ dispersion is close to $9.5$\,km\,s$^{-1}$. This is close to the adopted measurement error for such broad lines. However, the line profiles appear to vary significantly as confirmed by the TVS analysis. This is reminiscent of HD\,326331 \citep[see][]{san08} and we therefore conclude that the lines of HD\,46056 display some variations though they are not easily associated with the presence of a companion. For instance, non-radial pulsations could be responsible for changes in the line profiles of HD\,46056, likely to perturb the measurement of the radial velocities. More spectra and most of all a dedicated monitoring (spectroscopic as well as photometric) with a high temporal resolution are necessary to probe the existence of such short-time variations attributable to non-radial pulsations. The detection of such variations in the line profiles of rapid rotators has, for instance, been reported by \citet{deb08}.

To estimate the spectral classification, as the line profiles are clearly not Gaussian, we measured the EWs by integrating the line profiles. Conti's criteria yield a spectral type O8\,V. Moreover, the visual absolute magnitude of this object, calculated using $V\,=\,8.16$ and $E(B-V)\,=\,0.47$, is about $M_V\,=\,-4.25\,\pm$ 0.21 which corresponds to a MS star as suggested by the EWs of the classification lines. Unlike \citet{fro76} who noticed some sporadic emissions in the H${\alpha}$ line and assigned a spectral designation $e$ for this star, we report that the narrow emissions, visible in the Balmer H${\beta}$ and H${\alpha}$ lines, are probably caused by the presence of nebular material around HD\,46056. This conclusion is supported by the presence of other nebular lines, e.g. the [\ion{O}{iii}]\,$\lambda$\,5007 emission line. Therefore, a spectral type O8\,V\,n is assigned to HD\,46056, in agreement with \citet{wal73} or \citet{und90}.

\subsubsection{HD\,46202}

Although some authors claim that HD\,46202 may be RV variable \citep{abt72,und90}, we found no variation from our two year spectroscopic campaign (between January 2006 and January 2008) and we therefore conclude that HD\,46202 is a presumably single star. Indeed, the RV dispersions are similar to those measured for the interstellar lines and the average RVs agree with those reported in the literature. Our results therefore support the previous investigations made by \citet{pla31} and \citet{pet61}. 

The projected rotational velocity of HD\,46202, estimated using the Fourier method, amounts on average to $v\,\sin{i}\,=\,$54$\,\pm\,$15\,km\,s$^{-1}$ for the stellar lines.

Previous studies reported HD\,46202 as an O9\,V star \citep{mor55,mor65,wal71,con74,mas95}. Our data confirm this classification. Moreover, we obtained $M_V\,=\,-4.13\,\pm$ 0.21 from $V\,=\,8.18$ and $E(B-V)\,=\,0.44$, which indeed corresponds to a MS star.

 As quoted in the literature \citep{per91,mas95,mai04}, the visual magnitude seems constant in the range 8.18$-$8.21, with the exception of \citet{ogu81} who derived a V magnitude of 8.10.

\subsubsection{HD\,46223}

HD\,46223 is the hottest member of NGC\,2244. The star was regularly observed for studies of the interstellar medium \citep[e.g.][]{zag01} but the question of the variability of this star arose already in the 1970's. \citet{und90} concluded from published data \citep{cru74,con77,gie87} and from their own observations that the star may vary. The hypothesis of a spectroscopic binary was supported by the study of \citet{liu89} who noticed a RV dispersion of 10.3\,km\,s$^{-1}$ for three observations. However, these RV changes could also be interpreted as motions in the atmosphere of the star \citep{und90}.

The dispersions of the RVs of the more intense lines such as \ion{He}{ii}\,$\lambda$\,4542, \ion{He}{ii}\,$\lambda$\,4686 and \ion{He}{i}\,$\lambda$\,5876 are similar to those of the DIB. These values are slightly smaller than those of the metallic lines and the \ion{He}{i}\,$\lambda$\,4471 line. In any case, the dispersion is never significant even for these metallic lines: indeed, the larger noise of these faint lines might explain their slightly higher RV dispersion.

\begin{figure}[htbp]
  \centering
  \includegraphics[width=9cm, bb=27 202 590 710, clip]{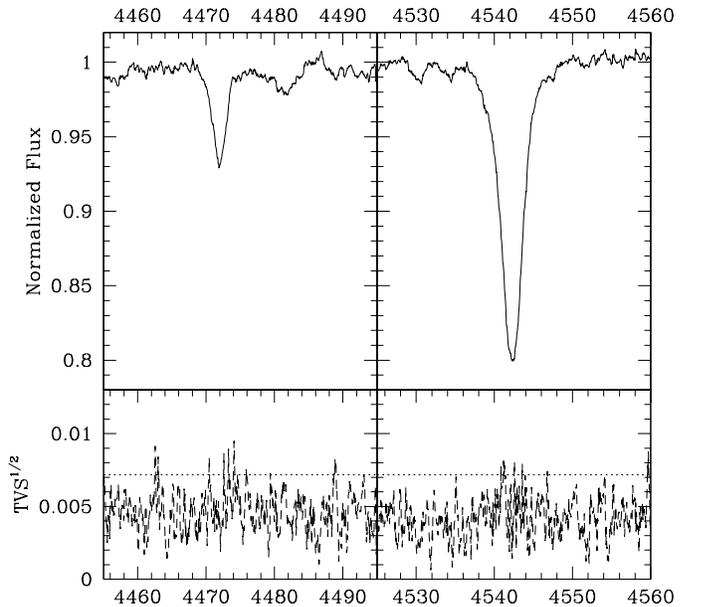}
  \caption{Mean spectrum and TVS of HD\,46223 computed from the Aurelie data for the \ion{He}{i}\,$\lambda$\,4471 (on the left) and the \ion{He}{ii}\,$\lambda$\,4542 lines (on the right). The dotted line illustrates the 99\% significance level for the variability evaluated following the approach of \citet{ful96}.}
  \label{Fig46223c}
\end{figure}

In consequence, we also applied the TVS method to our spectral time series. We performed this test with, on the one hand, only the Aurelie data, and, on the other hand, with all collected spectra (treated to get similar resolution, see Section \ref{section33}). In both cases, we failed to detect any significant variations around the \ion{He}{i}\,$\lambda$\,4471,\,$\lambda$\,5876, \ion{He}{ii}\,$\lambda$\,4542 and near the \ion{C}{iv}\,$\lambda \lambda$\,5801$-$12 lines (Fig.\,\ref{Fig46223c}) and we therefore consider the star to be constant. 

To determine the projected rotational velocity of the star, we used the Fourier transform on the same lines as for HD\,46150. We obtain a mean projected rotational velocity of about $v\,\sin{i}\,=\,$100$\,\pm\,$17\,km\,s$^{-1}$ while the literature reports a value of $v\,\sin{i}\,=\,$103\,km\,s$^{-1}$ \citep{pen96}.

HD\,46223 was previously classified as O5 \citep{hil56,joh62,bis82}, O4 \citep{mor65}, O5\,((f)) \citep{con74} and, finally, as O4\,V\,((f)) by \citet{mas95}. The measured EWs correspond to an O4 star. Using $V\,=\,7.27$ and $E(B-V)\,=\,0.50$, we compute a visual absolute magnitude of about $M_V\,=\,-5.23\,\pm$ 0.22, favoring a MS classification even though this value is slightly brighter than the theoretical one ($M_V$\hspace{-0.15cm}\textsuperscript{theo}$\,=\,-5.56$, \citealt{mar06}). Moreover, the spectrum clearly presents strong \ion{N}{iii}\,$\lambda \lambda$\,4634$-$41 emission lines accompanied by strong \ion{He}{ii}\,$\lambda$\,4686 absorption. HD\,46223 also displays both \ion{Si}{iv}\,$\lambda \lambda$\,4088 and 4116 in emission. Therefore, we assign an O4 V\,((f$^{+}$)) spectral type. We also note a weak emission in the H$\alpha$ line probably due to the nebular material surrounding the star as we already mentioned for HD\,46056.
 
\subsubsection{HD\,46485}

In previous works on NGC\,2244, little attention was paid to HD\,46485. The literature presents only a few measurements but in no way the results of a dedicated observing run. 

\begin{figure}[htbp]
  \centering
  \includegraphics[width=7cm, bb=110 156 430 696, clip]{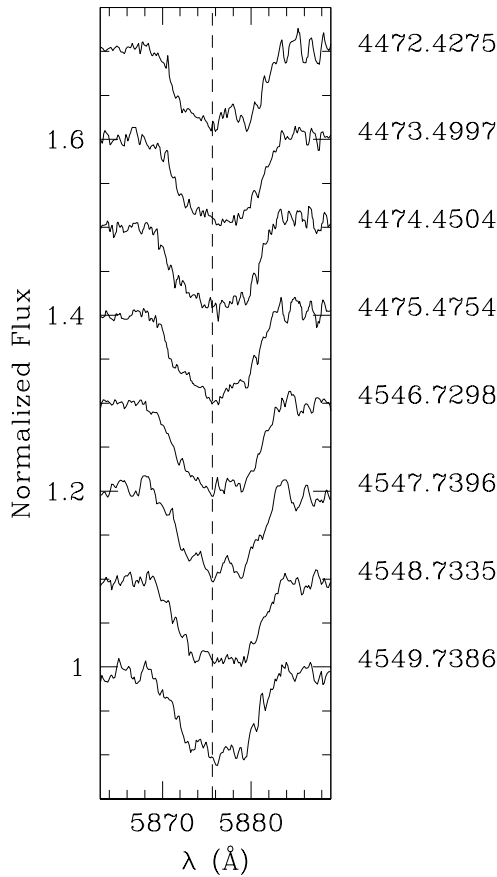}
  \caption{Same as Fig.\ref{Fig46056}, but for HD\,46485.}
  \label{Fig46485a}
\end{figure}

As for HD\,46056, HD\,46485 presents very broad and moderately deep lines suggesting that this star is a rapid rotator (Fig.\ref{Fig46485a}).

In order to determine the projected rotational velocity of HD\,46485, we independently applied the Fourier transform on three different lines: \ion{He}{ii}\,$\lambda$\,4686, \ion{O}{iii}\,$\lambda$\,5592, \ion{He}{i}\,$\lambda$\,5876 and we found a mean value of 301$\,\pm\,$25\,km\,s$^{-1}$. As the line profiles are not Gaussian, we used this measurement to generate several theoretical rotation profiles with $v\,\sin{i}$\,=\,325, 318, 291 and 301\,km\,s$^{-1}$ which we correlate with the observed \ion{He}{i}\,$\lambda$\,4471, \ion{He}{ii}\,$\lambda$\,4542, \ion{He}{ii}\,$\lambda$\,4686 and \ion{He}{i}\,$\lambda$\,5876 lines, respectively, to derive the RVs. The 1--$\sigma$ range of these lines are similar to each other (see Table \ref{tab2}) and, as for HD\,46056, similar to the RV measurement error. However, obvious profile changes are clearly detectable though they are difficult to attribute to a possible binarity (Fig.\,\ref{Fig46485a}). They may be due to non-radial pulsations.

\citet{con74} identified HD\,46485 as an O7.5 star while \citet{wal71} suggested an O8\,V\,n(e) classification. Using Conti's criteria, we infer a spectral type of O8 V for HD\,46485. With $V\,=\,8.20$ and $E(B-V)\,=\,0.58$, we derive a visual absolute magnitude $M_V\,=\,-4.68\,\pm$ 0.36. This value agrees with a typical O8\,V star \citep{mar06}. In consequence, with its broad diffuse lines, we classify HD\,46485 as an O8\,V\,n star.

\subsubsection{HD\,46966}

For our study about the variability of HD\,46966, we collected 22 spectra between January 2003 and January 2008. HD\,46966 does not present any significant RV dispersion neither for the metallic nor for the Helium lines ($\lambda$\,4471,\,$\lambda$\,4542,\,$\lambda$\,4686 and\,$\lambda$\,5876). Moreover, no significant variation is observed by the TVS method. This suggests that HD\,46966 is a presumably single star. 

\citet{pen96} estimated the projected rotational velocity of HD\,46966 close to 59\,km\,s$^{-1}$. Later on, \citet{mun99} revised the value to 90\,km\,s$^{-1}$. From four different lines (already mentioned for HD\,46573), we measured the $v\,\sin{i}$ with the Fourier method and the obtained mean value is estimated to be 63$\,\pm\,$17\,km\,s$^{-1}$ in good agreement with the value given by \citet{pen96}.

The spectral types quoted in the literature are O8\,V \citep{mun99,mai04} or O8.5\,V \citep{con77,gar80}. Based on the Conti criteria, we derive an O8.5 V spectral type. Although located outside NGC\,2244, we assume that HD\,46966 is situated approximately at the same distance as the open cluster (1.4--1.7 kpc). As before, we compute, with $V\,=\,6.87$ and $E(B-V)\,=\,0.21$, the minimum visual absolute magnitude $M_V\,=\,-4.51$ and the maximum one at about $M_V\,=\,-4.93$. These two values are larger than the theoretical one ($M_V$\hspace{-0.15cm}\textsuperscript{theo}$\,=\,-4.27$, \citealt{mar06}). However, we consider this difference as an effect of the uncertain distance rather than a consequence of binarity and we propose a MS classification for HD\,46966. 

\subsubsection{HD\,48279}

Previous spectroscopic investigations by \citet{pet61} reported a mean radial velocity of 31.0$\,\pm\,$5.1\,km\,s$^{-1}$. \citet{und90}, on the basis of their measured RV and also of those of \citet{pet61}, \citet{con77} and \citet{sm38}, suspected the star to be a single-lined spectroscopic binary.

We analyzed 4 spectra in the yellow domain and 10 in the blue region, along with 6 echelle spectra covering the whole visible domain, therefore including the blue and yellow wavebands (see Table \ref{table1}). The differences between RVs measured on Helium lines are not large enough to support the binary scenario. Although the variations are much stronger in the fainter metallic lines than in the stronger \ion{He}{i}\,$\lambda$\,4471 and\,$\lambda$\,5876 lines, the accuracy on the RV measurements of the more shallow lines is more affected by the noise. In addition, no signature of a companion, such as asymmetries in line profiles, has been found in any spectrum of our time series. The TVS method, first applied to spectra in the blue domain and then to the data in the yellow region, did not show any significant variations for \ion{He}{i}\,$\lambda$\,4471, \ion{He}{i}\,$\lambda$\,5876 nor for the metallic lines. The likelihood that HD\,48279 is a binary thus remains low. 

We estimated the projected rotational velocity from the Fourier method to $v\,\sin{i}\,=\,$130$\,\pm\,$13\,km\,s$^{-1}$. This value is in agreement with that of 123\,km\,s$^{-1}$ determined by \citet{coe77} and \citet{pen96}.

The spectrum of HD\,48279 presents, beside the Balmer lines, deeper \ion{He}{i} lines than \ion{He}{ii} ones, the obvious signature of a late O-type star and we infer an O7.5 classification from Conti's criteria. We also find \ion{Si}{iv}\,$\lambda \lambda$\,4088$-$4116 in deep absorption. The most remarkable features of the spectrum are the prominent  \ion{N}{iii} absorption lines at 4195, 4634 and 4641 \AA, implying an ON spectral designation.

\citet{ogu81} do not mention HD\,48279 as a member of NGC\,2244 but it still belongs to the Mon OB2 association \citep{dew04}. Without better information, we consider the star to be at the same distance as NGC\,2244. Using $V\,=\,7.91$ and $E(B-V)\,=\,0.407$, we compute the two extreme visual absolute magnitudes for a distance between 1.4 and 1.7 kpc. We obtain $M_V\,=\,-4.08$ and $M_V\,=\,-4.50$ respectively, situating HD\,48279 on the MS band. Accounting for the strength of the nitrogen lines, we hence adopt the ON7.5\,V spectral type for HD\,48279 slightly earlier than previous estimates (ON8 V, \citealt{con74}; O8\,V, \citealt{wal73}).

\subsection{ Photometry and close ``visual'' companions}

 In addition to our results, it is interesting to look at the photometric variability of each star. We thus retrieved, from the Hipparcos and the ASAS (All Sky Automated Survey) databases, the data corresponding to each studied object. These two datasets have been analyzed independently since the instrumentations are quite different. 

Table \ref{photo} lists the mean V magnitude of each star and the 1--$\sigma$ standard deviation (on V) of the best data provided for each star in the ASAS catalog. The Fourier transform applied to each studied object reveals no significant periodicity, concluding therefore that photometric variations of the stars are absent on the observed time series. 

\setcounter{table}{4}
\begin{table}[htbp]
\caption{The mean V magnitude and the 1--$\sigma$ standard deviation.}             
\label{photo}      
\centering          
\begin{tabular}{l c c}   
\hline\hline       
                      
Star & Mean V magnitude & 1--$\sigma$ error\\
\hline                    
HD\,46056 & 8.132 & 0.006\\
HD\,46149 & 7.598 & 0.006\\
HD\,46150 & 6.720 & 0.021\\
HD\,46202 & 7.946 & 0.048\\
HD\,46223 & 7.270 & 0.006\\
HD\,46485 & 8.255 & 0.011\\
HD\,46573 & 7.925 & 0.008\\
HD\,46966 & 6.854 & 0.019\\
HD\,48279 & 7.596 & 0.026\\
\hline
\hline
\end{tabular} 
\end{table}

The Hipparcos catalog provides us with a sufficient number of data for only four stars of our sample: HD\,46150, HD\,46485, HD\,46573 and HD\,46966. For these objects, we compute a mean magnitude and a corresponding 1--$\sigma$ standard deviation on the magnitude of 6.787$\,\pm\,$0.007, 8.342$\,\pm\,$0.023, 8.030$\,\pm\,$0.011 and 6.851$\,\pm\,$0.007, respectively. Once again, these small dispersions do not allow us to conclude to the existence of any significant variability for any of studied stars.

The speckle interferometric campaign of Galactic O-type stars with V$\,\lesssim\,$8 undertaken by \citet{mas98} revealed that HD\,46150 and HD\,48279 have ``visual'' companions whose separations range from 2.7$''$--74.6$''$ and 4.6$''$--56.2$''$, respectively. However, we note that the ``visual'' companion close to HD\,48279 was quoted as optical and not physical by \citet{lin85}. In addition, the Washington Visual Double Star (WDS, \citealt{mas01}) catalog also reported a ``visual'' companion located between 9.6$''$ and 10.5$''$ of HD\,46056. However we stress that no evidence exists so far to confirm with any certainty the physical association of these companions, especially with such high angular separations.

More recently, \citet{tur08} published results of an adaptative optic survey to search for faint companions among Galactic O-star systems (with V$\,\lesssim\,$8). They observed in the I-Band in order to detect companions in the projected separation between 0.5$''$ and 5.0$''$. They observed five stars of our sample (HD\,46149, HD\,46150, HD\,46223, HD\,46966 and HD\,48279) but found no companion.

\section{Discussion}

\subsection{Multiplicity of O-type stars}

 In this paper, we have studied all six O-stars of NGC\,2244: HD\,46056, HD\,46149, HD\,46150, HD\,46202, HD\,46223 and HD\,46485. HD\,46149 is clearly a long-period SB2 while HD\,46150 is a good binary candidate. The minimum frequency of spectroscopic binaries among O-stars in this cluster is thus at least 17\% and could possibly reach 33\%. 

\subsubsection{The detection biases}\label{4.1.1}

Because of the spectroscopic approach, a bias exists towards the detection of O$+$OB binary systems, providing only a lower limit on the true binary fraction. Indeed, several observational biases could make us miss a binary system with either a low orbital inclination, a rather large mass and/or brightness ratio or a very wide separation (and though either very long orbital period or a large eccentricity).

We first use the approach of \citet{gar80} to estimate the probability that we have missed a binary system because of a low orbital inclination as a function of the assumed mass ratio and orbital period. In this scheme, we consider that the semi-amplitude of the RV curve $K$ should be smaller or equal to twice the RV dispersion ($\sigma_{RV}$) of the photospheric lines, i.e. \ion{He}{ii}\,$\lambda$\,4542 in the blue domain and \ion{C}{iv}\,$\lambda$\,5812 in the yellow region (or \ion{He}{i}\,$\lambda$\,5876 when this metallic line is not visible). 
The mass function of the primary component is given by:
\begin{eqnarray}
f(m)&=&\frac{M_1~\sin^{3}i}{q~(1+q)^2}\\
&=&1.0355 \times 10^{-7}~K^3~P~(1-e^2)^{3/2}\label{eq2}
\end{eqnarray}
where $M_1$ is the primary mass (in $M_{\sun}$), $K$ is expressed in km\,s$^{-1}$, $e$ represents the orbital eccentricity, $q\,=\,M_1/M_2$ and $P$ defines the orbital period (in days).\\
Equation \ref{eq2} is transformed to express the orbital inclination as a function of other parameters. To obtain an upper limit on $\sin i$, we insert $2\sigma_{RV}$ as an upper limit on $K$ and assume a zero eccentricity.
Therefore, we obtain: 
\begin{equation}\label{eq3}
\sin i~\le~9.392~\times~10^{-3}~\sigma_{RV}~\left(\frac{P~q~(1+q)^2}{M_1}\right)^{1/3} 
\end{equation}
Making the assumption of a random distribution of the orbital directions in space, we can write the probability that the orbital inclination is smaller than the value obtained from Equation \ref{eq3} as:
\begin{equation}
\int_{0}^{i_{up}} \sin i~di~=~1-\cos i_{up}.
\end{equation}
The probabilities reported in Table \ref{tabprob}\footnote{Table \ref{tabprob} reports the probabilities that a binary system could have been missed for different values of the orbital period {\it P} and of the mass ratio {\it $q$}$\,=\,M_1/M_2$.} are the mean values computed from the two photospheric lines quoted above and by using as primary mass, the masses listed in \citet{mar05}. 

The probabilities that HD\,46202, HD\,46966 and HD\,48279 ($\simeq$\,20\,$M_{\sun}$) would be binaries with an orbital period less than 28\,days and a companion earlier than A5 ($\simeq$\,2\,$M_{\sun}$, $q$\,=\,10) are very low (i.e. $\le$ 7\%). Any missed binary would most probably have to be a system with a rather large mass ratio and a long orbital period.\\
The same conclusion could be reached for HD\,46223 ($\simeq$\,47\,$M_{\sun}$). At a period of 28 days, the probability of having an O-type companion is close to zero ($\sim$\,0.2\%) and we have only 3\% chance to miss a B3 secondary ($\simeq$\,8\,$M_{\sun}$, $q$\,=\,5). However, for the same orbital period, there is a 10\% chance that the binary signature could have been missed if the spectral type of the second component would be B6 or later ($\lesssim$\,5\,$M_{\sun}$, $q$\,=\,10).

\setcounter{table}{5}
\begin{table*}[htbp]
\caption{Probability to have missed a binary system.}
\label{tabprob}      
\centering          
\begin{tabular}{@{}l@{}|@{}c@{}|@{}c@{}|@{}c@{}}  
\hline  \hline 
& $P$\,=\,5 days  &$ P$\,=\,14 days  & $P$\,=\,28 days \\ 
\begin{tabular}{c}
\\
\hline
HD\,46056\\
HD\,46150\\ 
HD\,46202\\
HD\,46223\\
HD\,46485\\
HD\,46966\\
HD\,48279\\  
\end{tabular}
&
\begin{tabular}{c c c}
$q$=1&$q$=5&$q$=10\\
\hline
3.9E-3 & 5.0E-2 & 19.4E-2 \\ 
3.6E-3 & 4.6E-2 & 18.1E-2 \\
4.7E-4 & 6.0E-3 & 2.2E-2\\
6.4E-4 & 8.1E-3 & 2.9E-2\\
5.1E-3 & 6.7E-2 & 26.6E-2 \\
3.4E-4 & 4.3E-3 & 1.6E-2\\
2.3E-4 & 2.9E-3 & 1.0E-2\\
\end{tabular}
&
\begin{tabular}{c c c}
$q$=1&$q$=5&$q$=10\\
\hline
7.8E-3 & 10.3E-2 & 44.9E-2 \\ 
7.1E-3 & 9.5E-2  & 45.5E-2\\
9.4E-4 & 1.2E-2  &  4.3E-2 \\
 1.3E-3 & 1.6E-2 & 6.0E-2 \\
 1.0E-2 & 13.8E-2& 71.5E-1  \\
6.8E-4 & 8.6E-3  & 3.1E-2 \\
4.5E-4 & 5.7E-3  & 2.0E-2 \\
\end{tabular}
&
\begin{tabular}{c c c}
$q$=1&$q$=5&$q$=10\\
\hline
1.2E-2 & 16.9E-2 & \dots\\
1.1E-2 & 15.7E-2 & 34.7E-2\\
1.5E-3 & 1.9E-2 & 7.0E-2 \\
2.0E-3 & 2.6E-2 & 9.7E-2 \\
1.6E-2 & 23.1E-2 & \dots \\
1.1E-3 & 1.4E-2 & 5.0E-2 \\  
7.1E-4 & 9.0E-3 & 3.3E-2 \\  
\end{tabular}\\
\hline
\end{tabular} 
\begin{tabular}{@{}l@{}|@{}c@{}|@{}c@{}}    
\hline \hline
& $P$\,=\,60 days & $P$\,=\,300 days\\ 
\begin{tabular}{c}
\\
\hline
HD\,46056\\
HD\,46150\\ 
HD\,46202\\
HD\,46223\\
HD\,46485\\
HD\,46966\\
HD\,48279\\  
\end{tabular}
&
\begin{tabular}{c c c}
$q$=1&$q$=5&$q$=10\\
\hline
2.1E-2 & 30.4E-2 & \dots\\
1.9E-2 & 28.8E-2 & 78.2E-2\\
2.5E-3 & 3.2E-2 & 12.1E-2\\ 
3.4E-3 & 4.4E-2 & 17.2E-2\\
2.7E-2 & 43.3E-2 & \dots \\
1.8E-3 & 2.3E-2 & 8.5E-2 \\
1.2E-3 & 1.5E-2 & 5.5E-2 \\  
\end{tabular}
&
\begin{tabular}{c c c}
$q$=1&$q$=5&$q$=10\\
\hline
6.1E-2 & \dots & \dots\\
5.7E-2 & 53.3E-2 & \dots\\
7.3E-3 & 9.8E-2 & 51.5E-2\\ 
9.9E-3 & 13.7E-2 & 19.6E-2\\
8.2E-3 & \dots & \dots\\
5.3E-3 & 6.9E-2 & 29.0E-2\\
3.5E-3 & 4.5E-2 & 17.1E-2\\  
\end{tabular}\\
\hline
\end{tabular} 
\end{table*}

 We also perfomed Monte-Carlo simulations following the method of Sana et al. (2009, submitted). The orbital parameters are randomly generated (10000 trials) under the assumptions that 50\% of the binaries have an orbital period of $P$\,$\leq$\,10\,days while the other 50\% have $P$\,$>$\,10 days. Moreover, the periods are uniformly drawn in $\log\,P$ space (between 0.3 and 3.5), the system orientations ($\cos i$) are randomly drawn among a uniform distribution, the mass ratio $q$\,$=$\,$M_1/M_2$ is also uniformly distributed between 0.1 and 1.0 and we uniformly pick up eccentricities between 0.0 and 0.8. The simulations show that the probability (Table \ref{montecarlo}\footnote{Table \ref{montecarlo} lists theprobability of missing a significant orbital RV variation ($\Delta$RV\,$>$\,40\,km\,s$^{-1}$ for the two rapid rotators and $\Delta$RV\,$>$\,20\,km\,s$^{-1}$ for the other stars). $M_1$ represents the primary mass and is expressed in $M_{\sun}$. The last four columns are classified according to the orbital period of the system (in days).}) to have missed, in HD\,46202 and HD\,46223, a peak-to-peak RV variation ($\Delta$RV) larger than the observed value of 20\,km\,s$^{-1}$ is less than 1\% for systems with 2\,$\leq$\,$P$\,$\leq$\,10\,days and less than 10\% for 10\,$\leq$\,$P$\,$\leq$\,365\,days. For the rapid rotators, the simulations were run for $\Delta$RV$\,>\,$40\,km\,s$^{-1}$ since the measurement errors are larger. The probabilities amount to 2\% and 20\% for HD\,46056 and HD\,46485, respectively. Therefore, it is very unlikely that short-period O$+$OB systems were missed by our study, and this result does not depend on the small number of stars in NGC\,2244. For long-period (365--3000 days) systems, the probability of missing binaries increases to 50\% (70\% for rapid rotators). This is expected since our sampling does not permit us to investigate thoroughly this part of the parameter space.

\setcounter{table}{6}
\begin{table}[htbp]
\caption{Probability of missing a significant orbital RV variation.}             
\label{montecarlo}      
\centering          
\begin{tabular}{l c c c c c}   
\hline\hline       
                      
Star & $M_1$ & 2--10 & 10--365 & 365--3000 &2--3000\\
\hline                    
HD\,46056 & 20.8 & 0.02 & 0.17 & 0.77 & 0.24\\
HD\,46149 & 20.8 & 0.01 & 0.04 & 0.31 & 0.07\\
HD\,46150 & 34.4 & 0.01 & 0.03 & 0.21 & 0.05\\
HD\,46202 & 17.1 & 0.01 & 0.09 & 0.50 & 0.13\\
HD\,46223 & 46.9 & 0.01 & 0.03 & 0.22 & 0.05\\
HD\,46485 & 20.8 & 0.01 & 0.20 & 0.68 & 0.20 \\
HD\,46573 & 22.9 & 0.01 & 0.06 & 0.38 & 0.09\\
HD\,46966 & 18.8 & 0.01 & 0.06 & 0.34 & 0.08\\
HD\,48279 & 22.9 & 0.01 & 0.07 & 0.40 & 0.10\\
\hline
\hline
\end{tabular} 
\end{table}

\subsubsection{The effects of stellar evolution}

We now compare our results with those of other young open clusters studied in a similar way. The minimal binary fraction in NGC\,2244 (17\%) appears close to that of IC\,1805 (20\%). However, we note differences in the stellar content of these two clusters. \citet{deb06} have shown that at least one giant (O\,III\,(f)) and one supergiant (O\,If) belong to IC\,1805 while all O-type stars in NGC\,2244 are MS objects. Evolution effects (and thus the age of the systems) therefore seem to play no role in the binary properties though to be firmly established, this conclusion must be confirmed by studies of other clusters. 

 \citet{gar01} asserted that the orbital periods of spectroscopic binaries in the O-type star rich clusters were concentrated around 4--5 days. We emphasize that the absence of short-period binaries in NGC\,2244 does not support this conclusion (see previous subsection).\\ A comparison between NGC\,2244, IC\,1805 and NGC\,6231 (all three being now well-studied) is provided in Table\,\ref{tab4}. The first and second columns give the name and the density of the cluster. The third column provides the minimal binary fraction determined after intensive monitoring of the O-type stars. The next columns yield the number of O-stars in each cluster, the number of detected short and long-period binaries and the variable stars (potential binaries and probable intrinsic variables). The last column quotes relevant references. Differences are readily seen in this table. Apart from the already quoted difference in the proportion of short vs. long-period binaries, there is also a difference in the preferential nature of the companion. If we consider all O-type stars to be binaries, those apparently single having undetectable low-mass companions, the majority of O-stars in NGC\,6231 appears to have OB companions, while the signature of additional early-type objects is found in only a few O-stars of NGC\,2244 and IC\,1805.

\setcounter{table}{7}
\begin{table*}[htbp]
\caption{Summary of the multiplicity in young open clusters.}             
\label{tab4}      
\centering          
\begin{tabular}{c c c c c c c l}   
\hline\hline       
                      
Cluster & Density  & Minimal binary  & n\# O-stars & Short-period  & Long-period  & Variable  & References\\
 & (stars/pc$^2$) & fraction & & binaries (P\,$\le$\,10) & binaries & stars\\
\hline                    
NGC\,2244 & 2.84 & 17\% & 6 & 0 & 1 & 3 & This paper\\ \\
\hline
IC\,1805 & 1.65 & 20\% & 10 & 1 & 1 & 4$^{\star}$ & \citet{rau04} \\
& & & & & & & \citet{deb06}\\
 \hline                  
NGC\,6231 & 12.46 & 63\% & 16 & 6$^{\star \star}$  & 4 & 3 & \citet{san08} \\ \\
\hline
\end{tabular} 
\begin{list}{}{}
\item[$^{\star }$] Notes: Four late O-type stars were not studied by the two papers mentioned in the references. We will consider these objects as variable in the absence of a thorough study.
\item[$^{\star \star}$] Notes: We take into account that NGC\,6231 also contains a WR$+$O binary since a WR is an evolved O-star.
\end{list}
\end{table*}

\subsubsection{A possible impact of the stellar density}

The surface density of a given cluster is evaluated through the ratio between the number of members of the cluster and the circular area delineated by a typical radius, e.g. the one given by \citet{tad02}. Both IC\,1805 and NGC\,2244 have a rather loose cluster core whilst the stellar surface density appears much higher in the central part of NGC\,6231. With the new binary fraction from \citet{deb06}, \citet{san08} and this work, the anti-correlation between the cluster density and its massive star binary fraction proposed by \citet{gar01} can no longer be supported. On the contrary, a possible correlation might better reflect the behaviour detected in these three clusters. However, we must remain careful since this assumption is directly linked to the error bars on the binary fraction. If it is confirmed, this would emphasize the important role of dynamical gravitational interactions in the formation of massive binaries \citep{zin03}. To be ascertained, however, this relation needs to be confirmed by more data. The extreme case of loose OB associations might provide a good testbed to this aim.

\subsection{Rotational velocities}

HD\,46056 and HD\,46485 are two rapid rotators whose lines appear slightly variable. Rapid rotators could either
\begin{itemize}
\item[$\bullet$] be born as such and maintain a rapid rotation over their main-sequence lifetime,
\item[$\bullet$] be spun-up through transfer of mass and angular momentum during a Roche lobe overflow in a binary system,
\item[$\bullet$] be spun-up due to the core contraction at the end of the main-sequence.
\end{itemize}
For these two rapid rotators, it seems more likely that the first hypothesis is the correct one since we found no clear indication for multiplicity and there is no evidence that these objects are at or near the end of their main-sequence lifetime (chemical enrichment,...). 

\begin{figure}[ht!]
\begin{center}
\begin{minipage}{8cm}
\resizebox{8cm}{!}{\includegraphics[bb=40 180 580 675, clip]{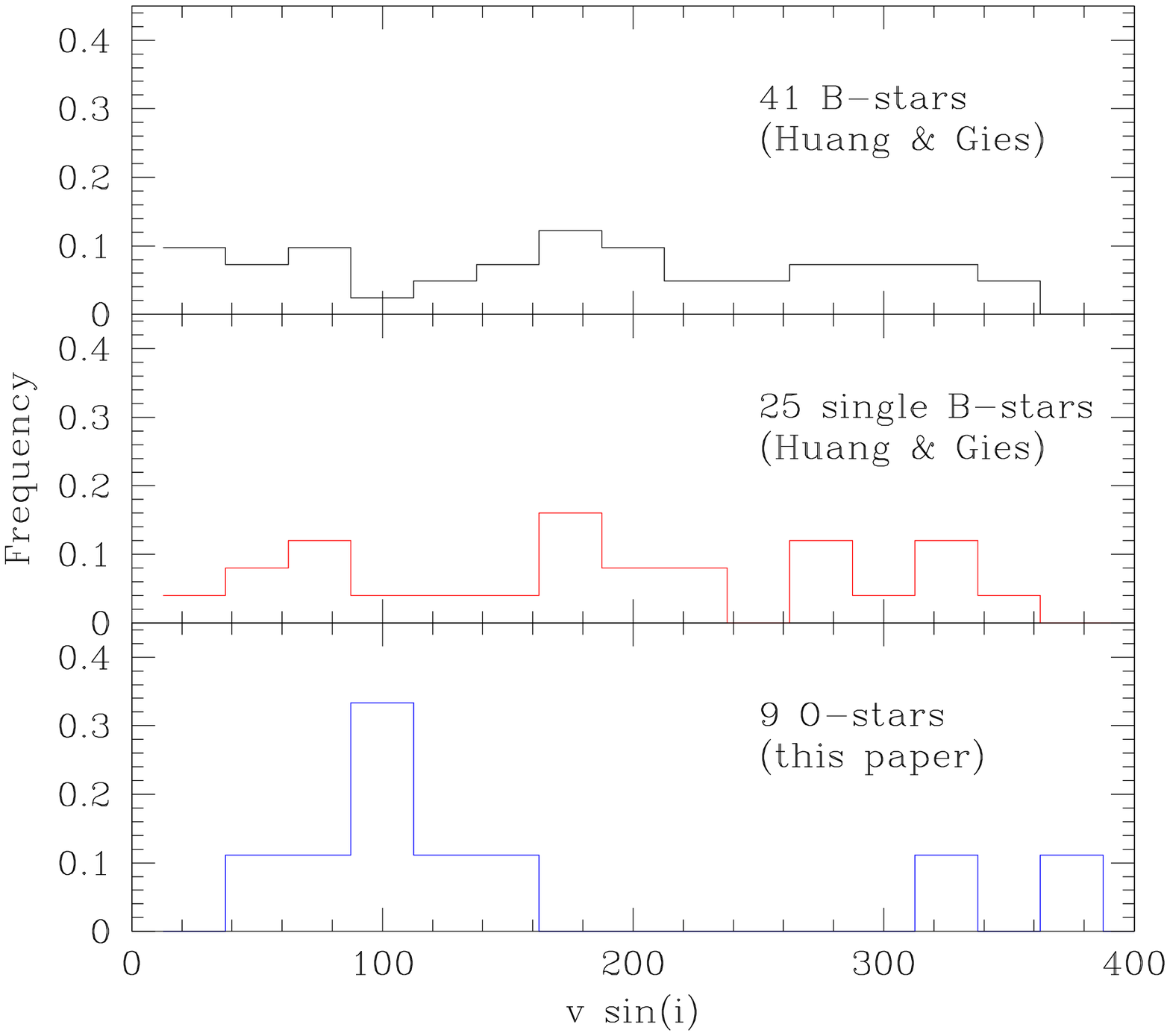}}
\end{minipage}
\hfill
\begin{minipage}{8cm}
\resizebox{8cm}{!}{\includegraphics[bb=47 225 560 600, clip]{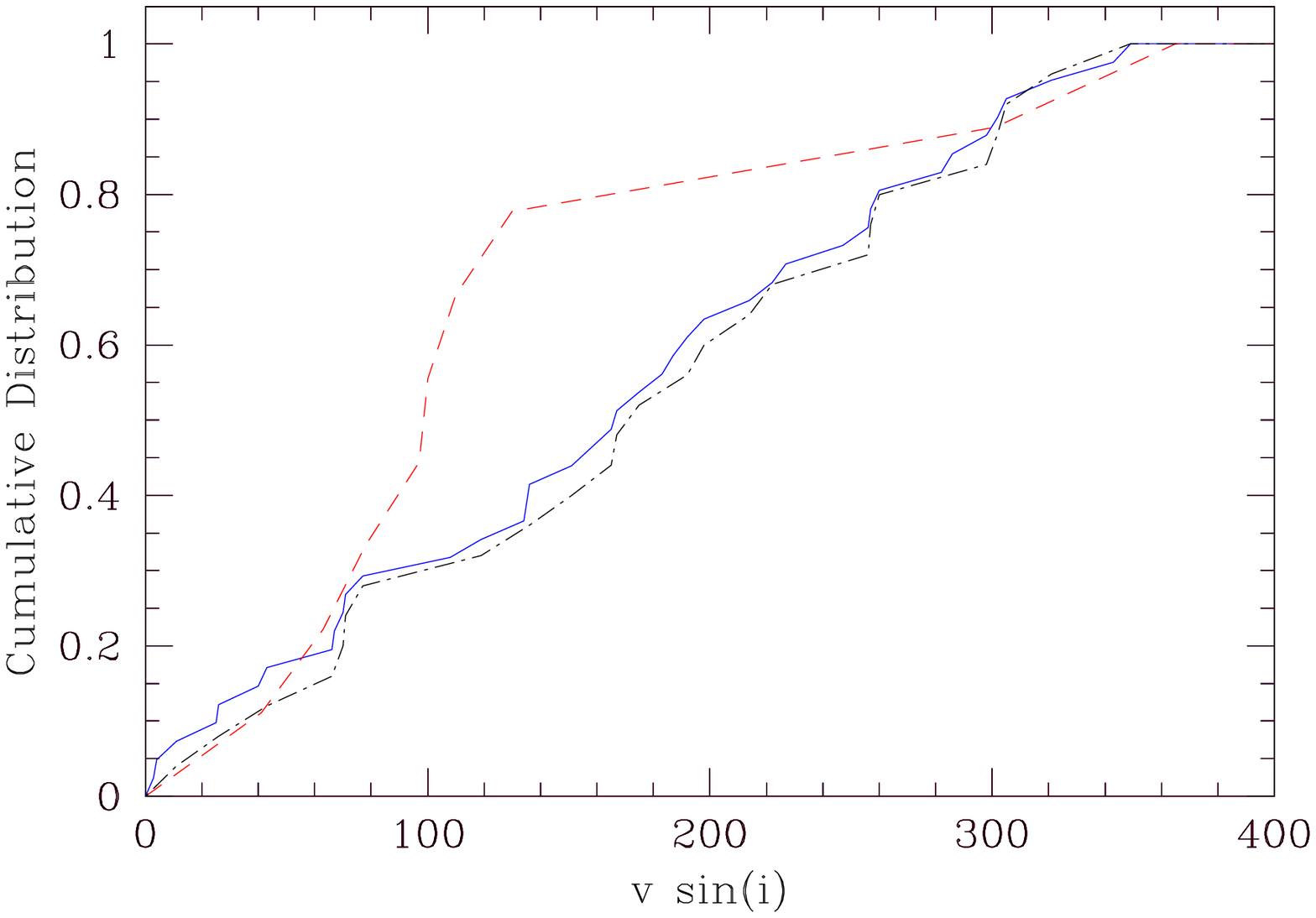}}
\end{minipage}
\caption{\label{histogram}{\it Top :} the distribution of projected rotational velocities of (1) the full sample of B stars studied by Huang \& Gies (2006, top panel), (2) those B stars of Huang \& Gies that show a constant radial velocity (middle panel) and (3) our sample of O-type stars. {\it Bottom :} the cumulative distributions of the three samples: all B-stars (solid line), single B stars (dashed-dotted line) and O-stars (dashed line).} 
\end{center}
\end{figure}

We also compared our distribution of the projected rotational velocities of the O-type stars in and around NGC\,2244 to the corresponding distribution for B-type stars. Indeed, \citet{hua06} performed a survey of projected rotational velocities for a large sample of mainly B-type (later than O9.5) main-sequence or giant stars in young open clusters, including NGC\,2244. The $v\,\sin{i}$ values were derived through comparison between theoretical line profiles and observed profiles of the He\,{\sc i}\,$\lambda \lambda$\,4026, 4387, 4471 and Mg\,{\sc ii}\,$\lambda$\,4481 lines. \citet{hua06} report projected rotational velocities of 41 stars in NGC\,2244. The average projected rotational velocity of the B-type stars amounts to 168\,km\,s$^{-1}$. The 15 stars that are earlier than B3 show an average $v\,\sin{i}$ of 153\,km\,s$^{-1}$. Five stars out of the total sample (i.e.\ 12\,\%) were found to display a projected rotational velocity above 300\,km\,s$^{-1}$. At first sight, the distribution of rotational velocities of the O-type stars looks different from that of the B-type stars (see Fig.\,\ref{histogram}). We have performed a two-sample Kolmogorov-Smirnov test to check whether or not the two distributions are significantly different. On the basis of this test and adopting a usual significance level of 0.05, the null hypothesis (both samples are drawn from the same parent distribution) cannot be rejected. This is true for both the full sample and the six O-stars of NGC\,2244.

\section{Summary and conclusions}

Our long-term monitoring has shed new light on the multiplicity of six massive O-stars in NGC\,2244 plus three in the association Mon OB2. Table \ref{tab3}\footnote{The first column provides the name of each star studied in this paper. The next two columns give the spectral classification and the projected rotational velocity (in\,km\,s$^{-1}$) of the O-type stars while the last column reports the status of the star as derived from our campaign: ``C'' means that our study did not reveal any indication of RV variability or of binarity, suggesting the star is probably single, ``Bin?'' indicates a potential binary, ``SB1'' corresponds to a spectroscopic binary where the second component has not been detected yet but where a period has been determined, whilst ``SB2'' corresponds to a spectroscopic binary where the two components are visible.} summarizes the present knowledge on these stars. 

In the association Mon OB2, the current study reveals the detection of a new SB1: HD\,46573. Although we determine a period of 10.67 days and we compute the orbital parameters of the system, these results must still be considered as preliminary. Indeed, the accuracy of the RV measurements could strongly influence the orbital solution of the system, considering the low amplitude of the radial velocity variations. The question of the eccentricity of the orbit remains open but we note that the eccentric radial velocity curve appears slightly better than the circular one.  

\setcounter{table}{8}
\begin{table}[htbp]
\caption{Summary of the optical properties of O-stars in NGC\,2244 (top part) and Mon OB2 (bottom).}             
\label{tab3}      
\centering          
\begin{tabular}{l c c c}   
\hline\hline       
                      
Name & Spectral Type & $v\,\sin{i}$ & Spectroscopic status\\ 
\hline                    
  HD\,46056 & O8Vn  & 355$\,\pm\,$21 & C ? (rapid rotator)\\
  HD\,46149 & O8V+B0--1V  & 78$\,\pm\,$11$^{\ddagger}$ & SB2\\
  HD\,46150 & O5.5V((f))  & 97$\,\pm\,$9 & Bin?\\
  HD\,46202 & O9V  & 54$\,\pm\,$15 & C\\
  HD\,46223 & O4V((f$^{+}$))  & 100$\,\pm\,$17 & C\\
  HD\,46485 & O8Vn  & 301$\,\pm\,$25 & C ? (rapid rotator)\\
\hline
  HD\,46573 & O7.5V((f)) & 110$\,\pm\,$18 & SB1\\
  HD\,46966 & O8.5V  & 63$\,\pm\,$17 & C\\
  HD\,48279 & ON7.5V  & 130$\,\pm\,$13 & C\\
\hline                  
\end{tabular} 
\begin{list}{}{}
\item[$^{\ddagger}$] Notes: The reported value for HD\,46149 is the projected rotational velocity for the primary.
\end{list}
\end{table}

Another interesting result, in NGC\,2244, is the detection, for the first time, of a secondary component in HD\,46149. However, for the time being, we can only constrain the period to be at least several tens of days. HD\,46149 is considered, by virtue of its mass and flux ratios, to be an O$+$B binary. An observation campaign should be initiated to monitor the star and better constrain the orbital solution of the system. 
Finally, we also detect weak variations in the RVs for one star belonging to NGC\,2244 (HD\,46150) and variations of the line profiles, most probably unrelated to binarity, in the two rapid rotators (HD\,46056 and HD\,46485). The observed binary fraction is thus 17\% at minimum.

This work, combined to the results of studies of NGC\,6231 
and IC\,1805, also revealed significant differences between clusters regarding the preference of O$+$OB systems and the distribution of short vs. long-period binaries. These differences can not be explained by the cluster age but could be linked to density effects. Of course, to confirm such a general conclusion, it is important to rely upon a large sample of well-studied very young open clusters that harbour a significant population of early-type stars. To date however, only a few clusters have been subject to such an intensive search for binary systems. It is therefore of paramount importance to perform similar studies on a larger sample of clusters, spanning as wide a range of physical properties as possible.

\begin{acknowledgements}
This work was supported by the FNRS (Belgium) and by a PRODEX XMM/Integral contract (Belspo) and by the Communaut\'e fran\c caise de Belgique - Action de recherche concert\'ee (ARC) - Acad\'emie Wallonie--Europe. The travels to OHP were supported by the Minist\`ere de l'Enseignement Sup\'erieur et de la Recherche de la Communaut\'e Fran\c caise. Philippe Eenens acknowledges support through CONACyT grant 67041. We also thank the staff of San Pedro M\`artir Observatory (Mexico) and of Observatoire de Haute-Provence (France) for their technical support. This work made use of the SIMBAD and the ASAS and Hipparcos databases.
\end{acknowledgements}

\clearpage
\onecolumn
\setcounter{table}{0}

\onllongtab{1}{
\begin{longtable}{l l l l c c c}
\caption{Observations used for the study of massive stars in both NGC\,2244 and Mon OB2.}\label{table1}\\
\hline\hline                 
Object & Cluster & Obs. run & Telescope/Instrument & Wavelength domain & n\# Spectra & $\Delta$T\\    
\hline                        
   HD\,46056 & NGC\,2244  &  Feb. 2006 & OHP - AURELIE/1.52m & 5500--5900\AA &6 &5.02\\      
    &   & Apr. 2007 & SPM - ESPRESSO/2.20m & &5&6.00\\
    &   & Nov. 2007 & OHP - AURELIE/1.52m & 4450--4900\AA & 1&\dots\\
    &   & Jan. 2008 & OHP - AURELIE/1.52m & 5500--5900\AA &4&3.00\\
    &   & Mar. 2008 & SPM - ESPRESSO/2.20m & &1&\dots\\
\hline
   HD\,46149 & NGC\,2244  &  Dec. 1994 & Asiago - Echelle/1.82m & &1&\dots\\
    &   & Dec. 1998 & Asiago - Echelle/1.82m & &1&\dots\\
    &   & Jan. 2003 & OHP - ELODIE/1.93m & &1&\dots\\
    &   & Nov. 2005 & OHP - ELODIE/1.93m & &1&\dots\\
    &   & Jan. 2006 & ESO - FEROS/2.20m & &1&\dots\\
    &   & Feb. 2006 & OHP - AURELIE/1.52m & 5500--5900\AA &6&5.10\\
    &   & Apr. 2007 & SPM - ESPRESSO/2.20m & & 3&3.97\\
    &   & Apr. 2007 & OHP - AURELIE/1.52m & 5500--5900\AA &1&\dots\\
    &   & Nov. 2007 & OHP - AURELIE/1.52m & 4450--4900\AA & 6&25.05\\
    &   & Jan. 2008 & OHP - AURELIE/1.52m & 5500--5900\AA &4&5.03\\
    &   & Mar. 2008 & SPM - ESPRESSO/2.20m & &5&4.93\\
    &   & Sep. 2008 & OHP - AURELIE/1.52m & 4450--4900\AA & 1&\dots\\
    &   & Oct. 2008 & OHP - AURELIE/1.52m & 4450--4900\AA & 3&4.02\\
\hline
   HD\,46150 & NGC\,2244  &  Nov. 1999 & OHP - ELODIE/1.93m & &1&\dots\\
    &   & Jan  2003 & OHP - ELODIE/1.93m & & 1&\dots\\
    &   & Nov  2004 & OHP - ELODIE/1.93m & & 1&\dots\\
    &   & Jan. 2006 & ESO - FEROS/2.20m & & 1&\dots\\
    &   & Feb. 2006 & ESO - FEROS/2.20m & & 1&\dots\\
    &   & Feb. 2006 & OHP - AURELIE/1.52m & 5500--5900\AA &6&5.10\\
    &   & Apr. 2007 & SPM - ESPRESSO/2.20m & & 5&5.99\\
    &   & Nov. 2007 & OHP - AURELIE/1.52m & 4450--4900\AA & 5&25.07\\
    &   & Jan. 2008 & OHP - AURELIE/1.52m & 5500--5900\AA &4&2.99\\
    &   & Mar. 2008 & SPM - ESPRESSO/2.20m & & 5&3.95\\
    &   & Oct. 2008 & OHP - AURELIE/1.52m & 4450--4900\AA & 4&4.00\\
\hline
   HD\,46202 & NGC\,2244  & Jan. 2006 & ESO - FEROS/2.20m & &1&\dots\\
    &   & Feb. 2006 & OHP - AURELIE/1.52m & 5500--5900\AA &2&1.98\\
    &   & Apr. 2007 & OHP - AURELIE/1.52m & 5500--5900\AA &2&5.00\\
    &   & Nov. 2007 & OHP - AURELIE/1.52m & 4450--4900\AA & 4&20.00\\
    &   & Jan. 2008 & OHP - AURELIE/1.52m & 5500--5900\AA &4&3.01\\
\hline
   HD\,46223 & NGC\,2244  & Jan. 1999 & OHP - ELODIE/1.93m & &1&\dots\\
    &   & Nov  2005 & OHP - ELODIE/1.93m & & 1&\dots\\
    &   & Jan. 2006 & ESO - FEROS/2.20m & & 1&\dots\\
    &   & Feb. 2006 & ESO - FEROS/2.20m & & 1&\dots\\
    &   & Feb. 2006 & OHP - AURELIE/1.52m & 5500--5900\AA & 6&5.03\\
    &   & Apr. 2007 & OHP - AURELIE/1.52m & 5500--5900\AA & 3&6.00\\
    &   & Nov. 2007 & OHP - AURELIE/1.52m & 4450--4900\AA & 5&21.98\\
    &   & Jan. 2008 & OHP - AURELIE/1.52m & 5500--5900\AA & 4&3.06\\
    &   & Mar. 2008 & SPM - ESPRESSO/2.20m & & 3&2.01\\
    &   & Oct. 2008 & OHP - AURELIE/1.52m & 4450--4900\AA & 1&\dots\\
\hline
   HD\,46485 & NGC\,2244  & Nov. 2005 & OHP - ELODIE/1.93m & & 1&\dots\\
    &   & Jan. 2006 & ESO - FEROS/2.20m & & 1&\dots\\
    &   & Nov. 2007 & OHP - AURELIE/1.52m & 4450--4900\AA & 6&14.99\\
    &   & Jan. 2008 & OHP - AURELIE/1.52m & 5500--5900\AA &4&3.05\\
    &   & Mar. 2008 & SPM - ESPRESSO/2.20m & &5&4.02\\
\hline
   HD\,46573 & Mon OB2  & Jan. 2006 & ESO - FEROS/2.20m & & 1&\dots\\
    &   & Feb. 2006 & ESO - FEROS/2.20m & & 1&\dots\\
    &   & Feb. 2006 & OHP - AURELIE/1.52m & 5500--5900\AA &3&2.97\\
    &   & Apr. 2007 & SPM - ESPRESSO/2.20m & & 5&5.99\\
    &   & Nov. 2007 & OHP - AURELIE/1.52m & 4450--4900\AA & 6&25.98\\
    &   & Jan. 2008 & OHP - AURELIE/1.52m & 5500--5900\AA &5&3.05\\
    &   & Oct. 2008 & OHP - AURELIE/1.52m & 4450--4900\AA & 3&3.00\\
\hline
   HD\,46966 & Mon OB2  & Jan. 2003 & OHP - ELODIE/1.93m & & 2&1.76\\
    &   & Nov  2004 & OHP - ELODIE/1.93m & &1&\dots\\
    &   & Jan. 2006 & ESO - FEROS/2.20m & &2&1.94\\
    &   & Feb. 2006 & OHP - AURELIE/1.52m & 5500--5900\AA &6&5.01\\
    &   & Feb. 2007 & OHP - AURELIE/1.52m & 5500--5900\AA &1&\dots\\
    &   & Apr. 2007 & OHP - AURELIE/1.52m & 5500--5900\AA &1&\dots\\
    &   & Nov. 2007 & OHP - AURELIE/1.52m & 4450--4900\AA & 5&24.05\\
    &   & Jan. 2008 & OHP - AURELIE/1.52m & 5500--5900\AA &4&3.02\\
\hline
   HD\,48279 & Mon OB2  & Jan. 2000 & OHP - ELODIE/1.93m & &1&\dots\\
    &   & Nov  2004 & OHP - ELODIE/1.93m & &1&\dots\\
    &   & Nov. 2007 & OHP - AURELIE/1.52m & 4450--4900\AA & 9&16.02\\
    &   & Jan. 2008 & OHP - AURELIE/1.52m & 5500--5900\AA &4&3.01\\
    &   & Mar. 2008 & SPM - ESPRESSO/2.20m & &4&4.01\\
    &   & Oct. 2008 & OHP - AURELIE/1.52m & 4450--4900\AA & 1&\dots\\
\hline                                   
\end{longtable}
}


\begin{thebibliography}{}

\bibitem[Abt 
  \& Biggs(1972)]{abt72} Abt, H.~A., \& Biggs, E.~S.\ 1972, in ``Bibliography of stellar radial velocities'', New York: Latham Process Corp., 1972, 
  
\bibitem[Bisiacchi et 
al.(1982)]{bis82} Bisiacchi, G.~F., Lopez, J.~A., \& Firmani, C.\ 1982, \aap, 107, 252 

\bibitem[Chen et al.(2007)]{che07} Chen, L., de Grijs, R., 
\& Zhao, J.~L.\ 2007, \aj, 134, 1368 

\bibitem[Conti 
  \& Alschuler(1971)]{con71} Conti, P.~S., \& Alschuler, W.~R.\ 1971, \apj, 170, 325 
  
\bibitem[Conti(1973)]{con73} Conti, P.~S.\ 1973, \apj, 179, 
  161
  
\bibitem[Conti 
\& Leep(1974)]{con74} Conti, P.~S., \& Leep, E.~M.\ 1974, \apj, 193, 113

\bibitem[Conti et al.(1977)]{con77} Conti, P.~S., Leep, 
  E.~M., \& Lorre, J.~J.\ 1977, \apj, 214, 759 
  
\bibitem[Conti 
  \& Ebbets(1977)]{coe77} Conti, P.~S., \& Ebbets, D.\ 1977, \apj, 213, 438 
  
\bibitem[Cruz-Gonz{\'a}lez et al.(1974)]{cru74} 
Cruz-Gonz{\'a}lez, C., Recillas-Cruz, E., Costero, R., Peimbert, M., 
\& Torres-Peimbert, S.\ 1974, Revista Mexicana de Astronomia y Astrofisica, 1, 211 

\bibitem[De Becker et 
  al.(2006)]{deb06} De Becker, M., Rauw, G., Manfroid, J., \& Eenens, P.\ 2006, \aap, 456, 1121 

\bibitem[De Becker et al.(2008)]{deb08} De Becker, M., 
Linder, N., 
\& Rauw, G.\ 2008, Information Bulletin on Variable Stars, 5841, 1 

\bibitem[de Wit et 
al.(2004)]{dew04} de Wit, W.~J., Testi, L., Palla, F., Vanzi, L., \& Zinnecker, H.\ 2004, \aap, 425, 937 

\bibitem[Feldmeier et 
al.(1997)]{fel97} Feldmeier, A., Puls, J., \& Pauldrach, A.~W.~A.\ 1997, \aap, 322, 878 

\bibitem[Frost 
\& Conti(1976)]{fro76} Frost, S.~A., \& Conti, P.~S.\ 1976, in IAU Symp. 70: Be and Shell Stars, 139 

\bibitem[Fullerton(1990)]{ful90} Fullerton, A.~W.\ 1990, 
  Ph.D.~Thesis  
  
\bibitem[Fullerton et al.(1996)]{ful96} Fullerton, A.~W., 
Gies, D.~R., \& Bolton, C.~T.\ 1996, \apjs, 103, 475 

\bibitem[Garc{\'{\i}}a 
\& Mermilliod(2001)]{gar01} Garc{\'{\i}}a, B., \& Mermilliod, J.~C.\ 2001, \aap, 368, 122 

\bibitem[Garmany et al.(1980)]{gar80} Garmany, C.~D., Conti, 
  P.~S., \& Massey, P.\ 1980, \apj, 242, 1063 
  
\bibitem[Gies(1987)]{gie87} Gies, D.~R.\ 1987, \apjs, 64, 545

\bibitem[Gonz{\'a}lez 
\& Levato(2006)]{gl06} Gonz{\'a}lez, J.~F., \& Levato, H.\ 2006, \aap, 448, 283 

\bibitem[Gosset et al.(2001)]{gos01} Gosset, E., Royer, P., 
Rauw, G., Manfroid, J., \& Vreux, J.-M.\ 2001, \mnras, 327, 435 

\bibitem[Heck et 
al.(1985)]{hec85} Heck, A., Manfroid, J., \& Mersch, G.\ 1985, \aaps, 59, 63 

\bibitem[Hensberge et 
  al.(2000)]{hen00} Hensberge, H., Pavlovski, K., \& Verschueren, W.\ 2000, \aap, 358, 553 
  
\bibitem[Hiltner(1956)]{hil56} Hiltner, W.~A.\ 1956, \apjs, 
2, 389 

\bibitem[Huang 
\& Gies(2006)]{hua06} Huang, W., \& Gies, D.~R.\ 2006, \apj, 648, 591 

\bibitem[Johnson(1962)]{joh62} Johnson, H.~L.\ 1962, \apj, 
136, 1135 

\bibitem[Kharchenko et 
al.(2005)]{kha05} Kharchenko, N.~V., Piskunov, A.~E., R{\"o}ser, S., Schilbach, E., \& Scholz, R.-D.\ 2005, \aap, 440, 403 

\bibitem[Li 
\& Smith(2005)]{li05} Li, J.~Z., \& Smith, M.~D.\ 2005, \aap, 431, 925 

\bibitem[Linder et 
al.(2008)]{lin08} Linder, N., Rauw, G., Martins, F., Sana, H., De Becker, M., \& Gosset, E.\ 2008, \aap, 489, 713 

\bibitem[Lindroos(1985)]{lin85} Lindroos, K.~P.\ 1985, \aaps, 60, 183 

\bibitem[Liu et al.(1989)]{liu89} Liu, T., Janes, K.~A., 
  \& Bania, T.~M.\ 1989, \aj, 98, 626 
  
\bibitem[Ma{\'{\i}}z-Apell{\'a}niz et al.(2004)]{mai04} 
Ma{\'{\i}}z-Apell{\'a}niz, J., Walborn, N.~R., Galu{\'e}, H.~{\'A}., 
\& Wei, L.~H.\ 2004, \apjs, 151, 103 

\bibitem[Martins et 
  al.(2005)]{mar05} Martins, F., Schaerer, D., \& Hillier, D.~J.\ 2005, \aap, 436, 1049 
  
\bibitem[Martins 
  \& Plez(2006)]{mar06} Martins, F., \& Plez, B.\ 2006, \aap, 457, 637 
  
\bibitem[Mason et al.(1998)]{mas98} Mason, B.~D., Gies, 
D.~R., Hartkopf, W.~I., Bagnuolo, W.~G., Jr., ten Brummelaar, T., 
\& McAlister, H.~A.\ 1998, \aj, 115, 821 

\bibitem[Mason et al.(2001)]{mas01} Mason, B.~D., Wycoff, 
G.~L., Hartkopf, W.~I., Douglass, G.~G., 
\& Worley, C.~E.\ 2001, \aj, 122, 3466 

\bibitem[Massey et al.(1995)]{mas95} Massey, P., Johnson, 
K.~E., \& Degioia-Eastwood, K.\ 1995, \apj, 454, 151 

\bibitem[Mathys(1988)]{mat88} Mathys, G.\ 1988, \aaps, 76, 427 
  
\bibitem[Mathys(1989)]{mat89} Mathys, G.\ 1989, \aaps, 81, 237 
  
\bibitem[Morgan et al.(1955)]{mor55} Morgan, W.~W., Code, 
A.~D., \& Whitford, A.~E.\ 1955, \apjs, 2, 41 

\bibitem[Morgan et al.(1965)]{mor65} Morgan, W.~W., Hiltner, 
  W.~A., Neff, J.~S., Garrison, R., 
  \& Osterbrock, D.~E.\ 1965, \apj, 142, 974 
  
\bibitem[Munari 
\& Tomasella(1999)]{mun99} Munari, U., \& Tomasella, L.\ 1999, \aaps, 137, 521 

\bibitem[Ogura 
  \& Ishida(1981)]{ogu81} Ogura, K., \& Ishida, K.\ 1981, \pasj, 33, 149 
  
\bibitem[Penny(1996)]{pen96} Penny, L.~R.\ 1996, \apj, 463, 
737 

\bibitem[P{\'e}rez(1991)]{per91} P{\'e}rez, M.~R.\ 1991, 
  Revista Mexicana de Astronomia y Astrofisica, 22, 99 
  
\bibitem[Petrie 
\& Pearce(1961)]{pet61} Petrie, R.~M., \& Pearce, J.~A.\ 1961, Publications of the Dominion Astrophysical Observatory Victoria, 12, 1 

\bibitem[Plaskett 
\& Pearce(1931)]{pla31} Plaskett, J.~S., \& Pearce, J.~A.\ 1931, Publications of the Dominion Astrophysical Observatory Victoria, 5, 1 

\bibitem[Rauw 
  \& De Becker(2004)]{rau04} Rauw, G., \& De Becker, M.\ 2004, \aap, 421, 693 

\bibitem[Reed(2005)]{ree05} Reed, C.\ 2005, VizieR Online 
Data Catalog, 5125, 0 

\bibitem[Sana et al.(2006a)]{san06a} Sana, H., Gosset, E., 
\& Rauw, G.\ 2006a, \mnras, 371, 67 
 
\bibitem[Sana et al.(2006b)]{san06b} Sana, H., Rauw, G., 
Naz{\'e}, Y., Gosset, E., \& Vreux, J.-M.\ 2006b, \mnras, 372, 661 

\bibitem[Sana et al.(2008)]{san08} Sana, H., Gosset, E., 
  Naz{\'e}, Y., Rauw, G., \& Linder, N.\ 2008, \mnras, 386, 447 
  
\bibitem[Sanford 
\& Merrill(1938)]{sm38} Sanford, R.~F., \& Merrill, P.~W.\ 1938, \apj, 87, 517 

\bibitem[Sim{\'o}n-D{\'{\i}}az 
  \& Herrero(2007)]{sim07} Sim{\'o}n-D{\'{\i}}az, S., \& Herrero, A.\ 2007, \aap, 468, 1063 
  
\bibitem[Stickland(1996)]{sti96} Stickland, D.~J.\ 1996, The 
  Observatory, 116, 294 
  
\bibitem[Tadross et al.(2002)]{tad02} Tadross, A.~L., Werner, 
P., Osman, A., \& Marie, M.\ 2002, New Astronomy, 7, 553 

\bibitem[Turner(1976)]{tur76} Turner, D.~G.\ 1976, \apj, 210, 
65 

\bibitem[Turner et al.(2008)]{tur08} Turner, N.~H., ten 
Brummelaar, T.~A., Roberts, L.~C., Mason, B.~D., Hartkopf, W.~I., 
\& Gies, D.~R.\ 2008, \aj, 136, 554 

\bibitem[Underhill 
  \& Gilroy(1990)]{und90} Underhill, A.~B., \& Gilroy, K.~K.\ 1990, \apj, 364, 626 
  
\bibitem[Walborn(1971)]{wal71} Walborn, N.~R.\ 1971, \apjs, 
23, 257 
  
\bibitem[Walborn(1973)]{wal73} Walborn, N.~R.\ 1973, \aj, 78, 
  1067 
  
\bibitem[Walborn 
  \& Fitzpatrick(1990)]{wal90} Walborn, N.~R., \& Fitzpatrick, E.~L.\ 1990, \pasp, 102, 379 (Erratum: 102 1094)
  
\bibitem[Wang et al.(2008)]{wan08} Wang, J., Townsley, L.~K., 
  Feigelson, E.~D., Broos, P.~S., Getman, K.~V., Rom{\'a}n-Z{\'u}{\~n}iga, 
  C.~G., \& Lada, E.\ 2008, \apj, 675, 464 

\bibitem[Wolfe et al.(1967)]{wol67} Wolfe, R.~H., Jr., Horak, 
H.~G., 
\& Storer, N.~W.\ 1967, The machine computation of spectroscopic binary elements (Modern astrophysics.~A memorial to Otto Struve), 251 

\bibitem[Zagury(2001)]{zag01} Zagury, F.\ 2001, New 
Astronomy, 6, 403

\bibitem[Zinnecker(2003)]{zin03} Zinnecker, H.\ 2003, in IAU Symp. 212: A 
Massive Star Odyssey: From Main Sequence to Supernova, 80 

\bibitem[Zinnecker 
\& Yorke(2007)]{zy07} Zinnecker, H., \& Yorke, H.~W.\ 2007, \araa, 45, 481 
  
\end{thebibliography}
\end{document}